\documentclass{aastex631}
\usepackage{graphicx}	% Including figure files
\usepackage{amsmath}	% Advanced maths commands
\usepackage{amssymb}	% Extra maths symbols
\usepackage{subfigure}
\usepackage{enumitem}

\shorttitle{Polarization of non-axisymmetric structured jets}
\shortauthors{J.-D. Li et al.}
\graphicspath{{./}{figures/}}
\begin{document}

\title{Polarization of gamma-ray burst afterglows in the context of non-axisymmetric structured jets}

\author{Jin-Da Li}
\affiliation{Institute for Frontier in Astronomy and Astrophysics, Beijing Normal University, Beijing 102206, China}
\affiliation{School of Physics and Astronomy, Beijing Normal University, Beijing 100875, China}

\author{He Gao}
\affiliation{Institute for Frontier in Astronomy and Astrophysics, Beijing Normal University, Beijing 102206, China}
\affiliation{School of Physics and Astronomy, Beijing Normal University, Beijing 100875, China}

\correspondingauthor{He Gao}
\email{gaohe@bnu.edu.cn}

\author{Shunke Ai}
\affiliation{Department of Astronomy, School of Physics and Technology, Wuhan University, Wuhan 430072, China}
\affiliation{Niels Bohr International Academy and DARK, Niels Bohr Institute, University of Copenhagen, Blegdamsvej 17, 2100, Copenhagen, Denmark}

\author{Wei-Hua Lei}
\affiliation{Department of Astronomy, School of Physics, Huazhong University of Science and Technology, Wuhan, Hubei 430074, China}

%% Note that the \and command from previous versions of AASTeX is now
%% depreciated in this version as it is no longer necessary. AASTeX 
%% automatically takes care of all commas and "and"s between authors names.

%% AASTeX 6.31 has the new \collaboration and \nocollaboration commands to
%% provide the collaboration status of a group of authors. These commands 
%% can be used either before or after the list of corresponding authors. The
%% argument for \collaboration is the collaboration identifier. Authors are
%% encouraged to surround collaboration identifiers with ()s. The 
%% \nocollaboration command takes no argument and exists to indicate that
%% the nearby authors are not part of surrounding collaborations.

%% Mark off the abstract in the ``abstract'' environment. 
\begin{abstract}
As the most energetic explosion in the universe, gamma-ray bursts (GRBs) are usually believed to be generated by relativistic jets. Some mechanisms (e.g. internal non-uniform magnetic dissipation processes or the precession of the central engine) may generate asymmetric jet structures, which is characterized by multiple fluctuations in the light curve of afterglow. Since the jet's structure introduces asymmetry in radiation around the line of sight (LOS),  it is naturally expected that polarization will be observable. In this work, we reveal the polarization characteristics of gamma-ray burst afterglows with a non-axisymmetric structured jet.
%, divided into independent patches along the azimuth angle. 
Our results show that the afterglow signal generally exhibits polarization, with the degree and evolution influenced by the specific jet structure, observing frequency, and the line of sight (LOS). The polarization degree is notably higher when the LOS is outside the jet. This degree fluctuates over time as different regions of radiation alternate in their dominance, which is accompanied by the rotation of the polarization angle and further reflects the intricate nature of the jet. Regarding its evolution over frequency, the polarization degree displays significant fluctuations at spectral breaks, with the polarization angle possibly undergoing abrupt changes.
%Additionally, at spectral breaks, the polarization degree exhibit significant fluctuations characterized by changes in polarization angle. 
%At frequencies where both synchrotron and SSC radiation affect polarization the local minimum polarization degree is greater than zero also implies the existence of asymmetric structures.} 
These features may provide strong evidence for future identification of potential GRBs with asymmetric jet structures. 
\end{abstract} 

%% Keywords should appear after the \end{abstract} command. 
%% The AAS Journals now uses Unified Astronomy Thesaurus concepts:
%% https://astrothesaurus.org
%% You will be asked to selected these concepts during the submission process
%% but this old "keyword" functionality is maintained in case authors want
%% to include these concepts in their preprints.

%% From the front matter, we move on to the body of the paper.
%% Sections are demarcated by \section and \subsection, respectively.
%% Observe the use of the LaTeX \label
%% command after the \subsection to give a symbolic KEY to the
%% subsection for cross-referencing in a \ref command.
%% You can use LaTeX's \ref and \label commands to keep track of
%% cross-references to sections, equations, tables, and figures.
%% That way, if you change the order of any elements, LaTeX will
%% automatically renumber them.
%%
%% We recommend that authors also use the natbib \citep
%% and \citet commands to identify citations.  The citations are
%% tied to the reference list via symbolic KEYs. The KEY corresponds
%% to the KEY in the \bibitem in the reference list below. 

\section{Introduction} \label{sec:intro}

Gamma-ray bursts are considered to be a kind of catastrophic stellar scale event located at cosmological distances \citep{2018pgrb.book}. Its energetic radiation suggests catastrophic astrophysical events, typically related to the core collapse of massive stars \citep{Woosley1993APJ,Paczynski1998,
MacFadyen1999APJ,Woosley2006ARAA} or the mergers
of two compact stellar objects (neutron star–neutron star and neutron
star–black hole binaries) \citep{Paczynski1986APJL,Eichler1989Nature,Paczynski1991bGRB,Paczynski1991aACTAA,Narayan1992APJL,Abbott2017PhysRevLett}. It is well established that the GRBs are driven by a ultrarelativistic jet from a central engine (neutron stars or black holes). The dissipation of magnetic or kinetic energy in the jet drives the prompt emission a GRB \citep{Rees1994ApJ,MeszarosRees1997ApJ,ZhangB2011ApJ}, and the interaction between the jet and the interstellar medium provides a long-lasting multi-wavelength 
afterglow emission \cite[][for a review]{Gao2013MNRAS}.  
%the later emission, afterglow, is \textbf{It is believed that the forward shock (FS), generated by the interaction between the jet and the interstellar medium, accelerates electrons in the environment. Both the synchrotron radiation from these accelerated electrons and the Synchrotron Self-Compton (SSC) scattering radiation contribute to the long-term afterglow, spanning from radio to TeV. During these processes, the magnetic field plays a crucial role in the radiation mechanism. 
The main radiation mechanisms that produce the afterglow signal are synchrotron radiation (which dominates the radiation from radio to X-ray bands) and synchrotron self-Compton scattering radiation (which dominates the radiation in the high-energy gamma-ray band). Since both of these radiation mechanisms inherently produce highly polarized signals, there has been a long history of research on the polarization of gamma-ray burst afterglows.

\citet{Medvedev1999ApJ} have discussed a process of magnetic field generation and geometric structure, suggesting that the magnetic field is completely tangled in the front plane of shock, but has high coherence in the orthogonal direction. In this case, polarization can only be observed when the emission regions is asymmetrical relative to the observer's line of sight (LOS) \citep{Ghisellini1999MNRAS}. The polarization properties have been studied when the LOS is not observed along the symmetry axis of the jet \citep{Sari1999ApJ,Ghisellini1999MNRAS,Rossi2004MNRAS}. Here one assumed that the angle between the LOS and the axis of the jet is $\theta_{\text{obs}}$, the half angle of the jet is $\theta_{\text{j}}$, and the Lorentz factor of the jet is $\gamma$. As long as $1/\gamma>\theta_{\text{j}}-\theta_{\text{obs}}$, the emission region is asymmetric for the observer. In this case, the degree of polarization can achieve $10\%$ approximately, and the polarization direction can only be parallel or perpendicular to the plane where the LOS and jet axis are located \citep{Ghisellini1999MNRAS}. It should be noted that in this case, the equal arrival time surface (EATS) effect may affect the polarization results, especially in the case of off-axis observations \citep{Huang2007ChJAA,Lan2023ApJL}.

%However, early research on polarization often overlooked the equal arrival time surface (EATS) effect, which might be important for observation, especially off-axis detection \citep{Huang2007ChJAA}. \citet{Lan2023ApJL} investigate the afterglow polarization with the EATS effect, and found that the polarization degree bump move to later times as the EATS effect increases, which is particularly important in off-axis observations.

%If the magnetic field is completely tangled over the entire range of emission regions seen by the observer, the generated radiation is unpolarized. This means that the polarization of each point source observed by the observers can be completely counterbalance. On the other hand, polarization may be observed when the emission regions is asymmetrical relative to the observer's line of sight (LOS). Therefore, there may be some degree of polarization that can be observed in the afterglow of GRBs. Conceivably, the evolution of polarization can provide crucial insights into the properties of the jets and central engines.

The asymmetric distribution of the radiation region around the LOS caused by the jet structure can also result in obvious polarization characteristics of the afterglow signal.
%Similarly, the asymmetry of the emission region around the LOS caused by the jet structure results in polarization of the afterglow. 
\citet{Rossi2004MNRAS} have discussed the polarization properties of power-law and Gaussian jets. They found that the evolution of polarization is sensitive to the luminosity distribution of the jet. For power-law jets, the time at which polarization peaks aligns with the break of light curve, which contrasts with the moment of minimum polarization observed at this moment in homogeneous jets. On the other hand, the exponential wings of the Gaussian jets shift the peak time of polarization after the break in the light curve. \citet{Wu2005MNRAS} studied the polarization of two-component jets. Its polarization evolution largely depends on the ratio of the intrinsic parameters of the two components, lateral expansion and observation angle. Over a broad range of viewing angle, the narrow component exert a dominant influence on polarization for an extended duration. The common feature of these structures is that they are symmetrical relative to the jet axis, and the possible polarization angle is the same as that of homogeneous jets.

Except for symmetrical structure, some physical processes may generate asymmetric structured jets, e.g. non-uniform magnetic dissipation within the jet, as suggested by \citet{Narayan2009MNRAS} and the precession of central engines in GRBs, as indicated by \citet{Huang2019MNRAS}. More recent work by \citet{Lamb2022Univ} explored 3D hydrodynamic jets in neutron star mergers environment and indeed found some rotational variation of inhomogeneity. Under the framework of asymmetric structures, some models were proposed to explain the prompt emission properties of GRBs, such as the jet hotspots, the patchy shells and the micro/sub jets \citep{Nakamura2000ApJL,Yamazaki2004APJL,Ioka2005ApJ}. On the other hand, \citet{Meszaros1998APJ} noted differences in afterglow due to the angular anisotropy of fireballs. Most recently, \citet{Li2023MNRAS} conducted a detailed analysis of the potential characteristics of GRB afterglows within the framework of non-axisymmetric structured jets, 
%where the jet is divided into individual elements as a step function of the azimuth angle, 
and they found that radiative contributions from multiple elements may lead to the appearance of multiple distinct peaks or plateaus in the light curve. 

Here we will delve into the polarization properties of the afterglow signal generated by the non-axisymmetric jet. We will present the evolution of the degree of polarization and the polarization angle over time at a given observation frequency. Additionally, we will also demonstrate how the degree of polarization and the polarization angle vary with frequency at a given time. Lastly, we will thoroughly examine the impact of different LOS directions on the results.

\section{Model Description} 
\label{math}

\begin{figure}[htbp]
    \centering
    \includegraphics[width=8.5cm]{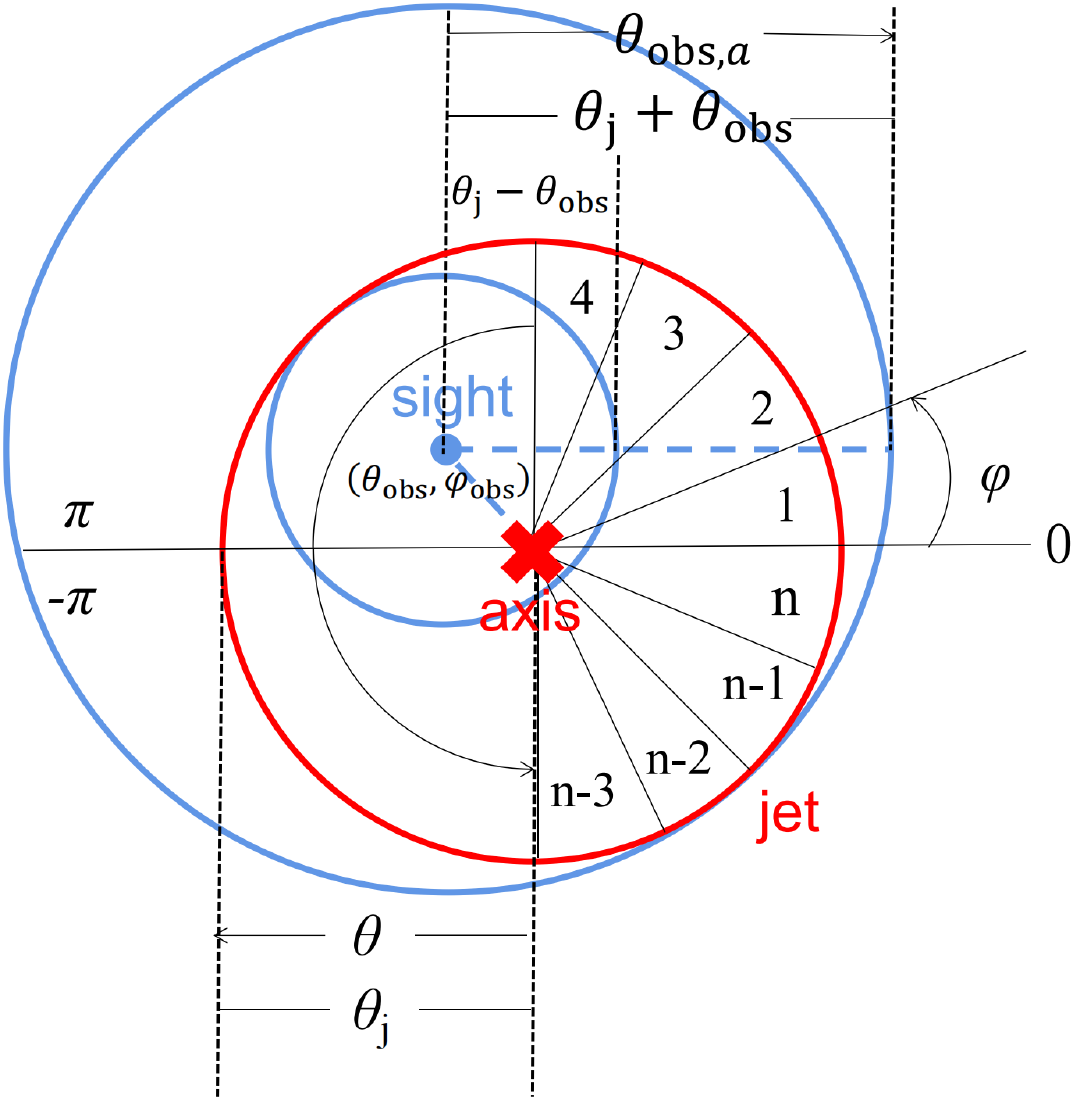}
    \caption{
    The diagram illustrates the jet structure and coordinate system. The red circle represents the cross-section of a jet with a half opening angle of $\theta_{\text{j}}$. Using the jet axis as the coordinate original point, from axis ($\theta=0$) to the edge of the jet ($\theta=\theta_{\text{j}}$) is the $\theta$ direction, and the circumference is $\varphi$ direction with $\varphi\in\left[-\pi,\pi\right]$. The distribution of physical parameters on the jet circumference are step function of $\varphi$, and the cross-section is divided into $N$ elements. The projection of the observer's LOS on the jet cross-section is marked as "sight", and the coordinate is $\left(\theta_{\text{obs}},\varphi_{\text{obs}}\right)$. The emission of the jet is the integral of the circles with the radius ($\theta_{\text{obs},a}$) from 0 to $\theta_{\text{j}}+\theta_{\text{obs}}$ around the LOS. For a uniform jet, only the emission between $\theta_{\text{j}}-\theta_{\text{obs}}$ and $\theta_{\text{j}}+\theta_{\text{obs}}$ is polarized.}
    \label{coordinate}
\end{figure}

\citet{Li2023MNRAS} have established an asymmetric jet modeling method by dividing the jet into $N$ elements and analyzed the afterglow light curve. One can describe the asymmetric jet on the cross-section of the jet using polar coordinates (see Figure \ref{coordinate}). The jet axis is set as the coordinate original point, from axis ($\theta=0$) to the edge of the jet ($\theta=\theta_{\text{j}}$) is the $\theta$ direction, and the circumference is $\varphi$ direction with $\varphi\in\left[-\pi,\pi\right]$. The $N$ elements represent the distribution of parameters are step function of $\varphi$. Each element can be approximated as a uniform "patch". The angle between LOS and jet axis is $\theta_{\text{obs}}$. And the projection's polar angle of the LOS on the jet cross-section is $\varphi_{\text{obs}}$.

The ejecta of the jet is assumed as a magnetized plasma slab. 
%If observed perpendicularly to the slab, the magnetic field structure is completely tangled. And if observed parallelly to the slab, the magnetic field structure is to some extent ordered. 
When viewed perpendicular to the slab, the magnetic field structure appears to be completely tangled. Conversely, when observed parallel to the slab, the magnetic field structure exhibits a certain degree of ordering. Such a magnetic field can be generated through unidirectional compression of a 3-dimensional tangled magnetic field \citep{Laing1980MNRAS} or Weibel instability \citep{Medvedev1999ApJ}. Assuming the degree of polarization observed parallel to the slab is $P_0$, at an angle $\rho'$ from the normal of the slab, the polarization can be expressed as \citep{Laing1980MNRAS}:
\begin{equation}
P\left(\rho'\right)=\frac{\text{sin}^2\rho'}{1+\text{cos}^2\rho'}P_0.
\end{equation}
The value of $P_0$ is determined by the mechanism that produces the photons. For instance, we consider both synchrotron radiation and synchrotron self-Compton scattering in this work. $P_0$ for synchrotron photons depends on the index of the power-law distribution of the radiating electrons and can be calculated by averaging over their isotropic distribution \citep{Longair1994}. $P_0$ for photons originating from SSC scattering is assumed to be $100\%$ \citep{Gill2020MNRAS}.
%For synchrotron radiation, $P_0$ is related to the power law distribution of electrons and can be calculated by averaging over their isotropic distribution \citep{Longair1994}. On the other hand, for synchrotron self-Compton (SSC) scattering, the $P_0$ is almost $100\%$ \citep{Gill2020MNRAS}.}
Since the jet moves in the direction perpendicular to the slab with Lorentz factor $\gamma$, the relativistic aberration of photons cannot be ignored. The photons observed at $\rho=1/\gamma$ in observer's frame correspond to those at $\rho'=\pi/2$ in jet's comoving frame.

For the $i-$th element, the polarization of each point source on an element with an angle $\theta_{\text{obs},a}$ from the LOS can be written in the complex number field as $\vec{P}=P\left(\theta_{\text{obs},a}'\right)e^{2i\theta_{\text{p}}}$ \citep{Ghisellini1999MNRAS}, where $\theta_{\text{p}}$ is the polarization angle of the linear polarization in the observer frame. If the position angle relative to the LOS is set to be $\chi$, then we have $\theta_{\text{p}}=\chi$ for synchrotron radiation \citep{Ghisellini1999MNRAS} and $\theta_{\text{p}}=\chi+\pi/2$ for SSC scattering \citep{Gill2020MNRAS}. All point sources with equal $\theta_{\text{obs},a}$ form a loop around the LOS (see Figure \ref{coordinate}). The polarization of the jet at an angle $\theta_{\text{obs},a}$ from the LOS results from the combined contributions of each point source arranged in a ring, which can be expressed as an integral with respect to $\chi$. But only the portion of each loop within the element contributes to the polarization. If the position angle range of the loop within the element is defined as from $\chi_1$ to $\chi_2$, the polarization of the loop is
\begin{equation}
\vec{P}_i\left(\theta_{\text{obs},a}\right)=P_i\left(\theta_{\text{obs},a}'\right)\int_{\chi_1}^{\chi_2}e^{2i\theta_{\text{p}}}\text{d}\chi.
\end{equation}
The polarization of the $i-$th element is the integral over the $\theta_{\text{obs},a}$:
\begin{equation}
\vec{P}_i=\frac{1}{F_{\nu,i}}\int_{0}^{\theta_{\text{j}}+\theta_{\text{obs}}}F_{\nu,i,a}\left(\theta_{\text{obs},a}\right)\vec{P}_i\left(\theta_{\text{obs},a}\right)\text{d}\theta_{\text{obs},a}.
\end{equation}
And the polarization of the jet is:
\begin{equation}
\vec{P}_t=\frac{1}{F_{\nu,t}}\int_{0}^{\theta_{\text{j}}+\theta_{\text{obs}}}\sum\limits_{i=1}^{n}F_{\nu,i,a}\left(\theta_{\text{obs},a}\right)\vec{P}_i\left(\theta_{\text{obs},a}\right)\text{d}\theta_{\text{obs},a}.
\end{equation}

The $F_{\nu,t}$ is the total radiation flux of afterglow of the asymmetric jet with $N$ elements, which is the sum of afterglow flux from each element:
\begin{equation}
F_{\nu,t}=\sum\limits_{i=1}^{n}F_{\nu,i}.
\end{equation}
The $F_{\nu,i,a}\left(\theta_{\text{obs},a}\right)$ is the afterglow radiation flux of a loop with radius $\theta_{\text{obs},a}$. And the afterglow radiation flux of the $i-$th element $F_{\nu,i}$ is the integral of $F_{\nu,i,a}$ over $\theta_{\text{obs},a}$:
\begin{equation}
\begin{split}
F_{\nu,i}&=\int_{0}^{\theta_{\text{j}}+\theta_{\text{obs}}}F_{\nu,i,a}\left(\theta_{\text{obs},a}\right)\text{d}\theta_{\text{obs},a}\\
&=\int_{0}^{\theta_{\text{j}}+\theta_{\text{obs}}}F_{\nu,a}\left(\theta_{\text{obs},a}\right)\chi_{i}\left(\theta_{\text{obs},a}\right)\text{d}\theta_{\text{obs},a}.
\end{split}
\end{equation}
Where $\chi_{i}\left(\theta_{\text{obs},a}\right)$ is the total position angle of the loop with $\theta_{\text{obs},a}$ in the $i-$th element, which is the part on a loop that contributes to the radiation of afterglow. And $F_{\nu,a}$ represents the afterglow radiation flux of a point source on the element with an angle $\theta_{\text{obs},a}$ to LOS. It worth noting that \citet{Lan2023ApJL} found the influence of the EATS effect on polarization, the EATS effect shouldn't be ignored. The radiation flux needs to be transferred to the observer direction through Doppler conversion. Therefore, $F_{\nu,a}$ is represented by \citep{Granot2002ApJL}
\begin{equation}
F_{\nu,a}=a^3F_{\nu/a}\left(at\right),
   \label{lzf}
\end{equation}
with a factor
\begin{equation}
    a=\frac{1-\beta}{1-\beta\cos{\theta_{\text{obs},a}}}\approx\frac{1}{1+\gamma^2\theta_{\text{obs},a}^2},
    \label{lz}
\end{equation}
where $F_{\nu}\left(t\right)$ is the flux at $\theta_{\text{obs},a}=0$ along the observer time $t$ of the point source. It is related to the physical properties of the element and the electrons in the interstellar medium. The formulae for calculating the specific flux $F_{\nu}$ for synchrotron radiation and SSC scattering are detailed in Appendix.

\section{Polarization properties for a non-axisymmetric jet}

With the formula introduced in Section \ref{math}, here we show the polarization properties for a non-axisymmetric jet in some specific cases. 
%here we calculate the polarization evolution \textbf{and spectral distribution} of the afterglows in some typical cases
In this section, we only present the results of jets decelerating in uniform interstellar medium.

\subsection{Asymmetric Jet With 2 Elements}

\begin{figure}[htbp]
    \centering
    \includegraphics[width=8.5cm]{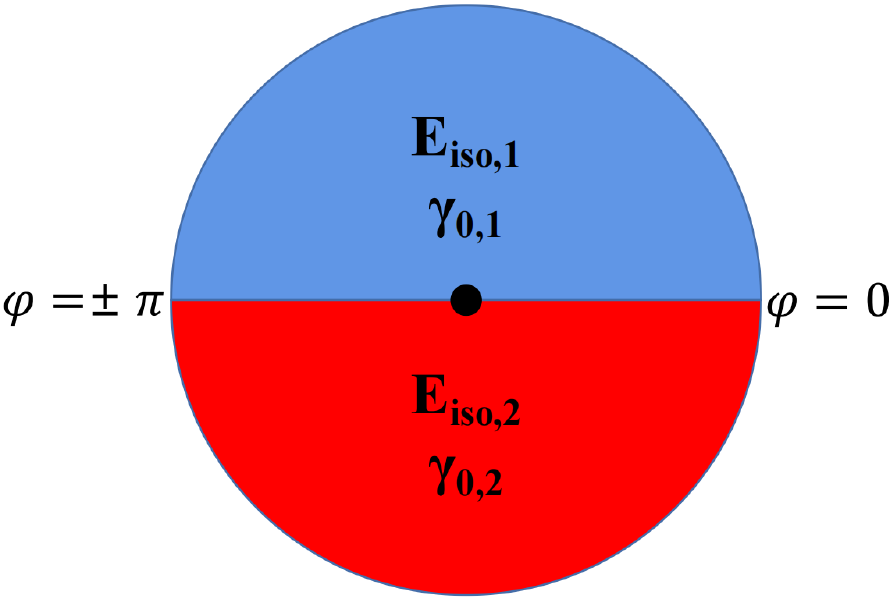}
    \caption{
    Schematic diagram of a asymmetric jet's cross-section with 2 elements. The interface between 2 elements local at $\varphi=0$ or $\varphi=\pm\pi$.}
    \label{3.1}
\end{figure}

The simplest asymmetric jet structure is characterized by a sharp interface between two distinct elements resulting from variations of the physical parameters $\gamma_{0}$ and $E_{\text{iso}}$ at different azimuth $\varphi$. The interface between two elements is defined as $\varphi=0$ or $\varphi=\pm\pi$. Following \citet{Li2023MNRAS}, we describe this structure as:
\begin{equation}
    \gamma_0=\begin{cases}
    \gamma_{0,1}&0<\varphi<\pi,\\
    \gamma_{0,2}&\text{others},
    \end{cases}
    \label{g0}
\end{equation}
and
\begin{equation}
    E_{\text{iso}}=\begin{cases}
    E_{\text{iso},1}&0<\varphi<\pi,\\
    E_{\text{iso},2}&\text{others}.
    \end{cases}
    \label{E0}
\end{equation}
Figure \ref{3.1} shows the cross section for the asymmetric jet.

\begin{figure*}[htbp]
    \centering
    \subfigure{\includegraphics[width=17cm]{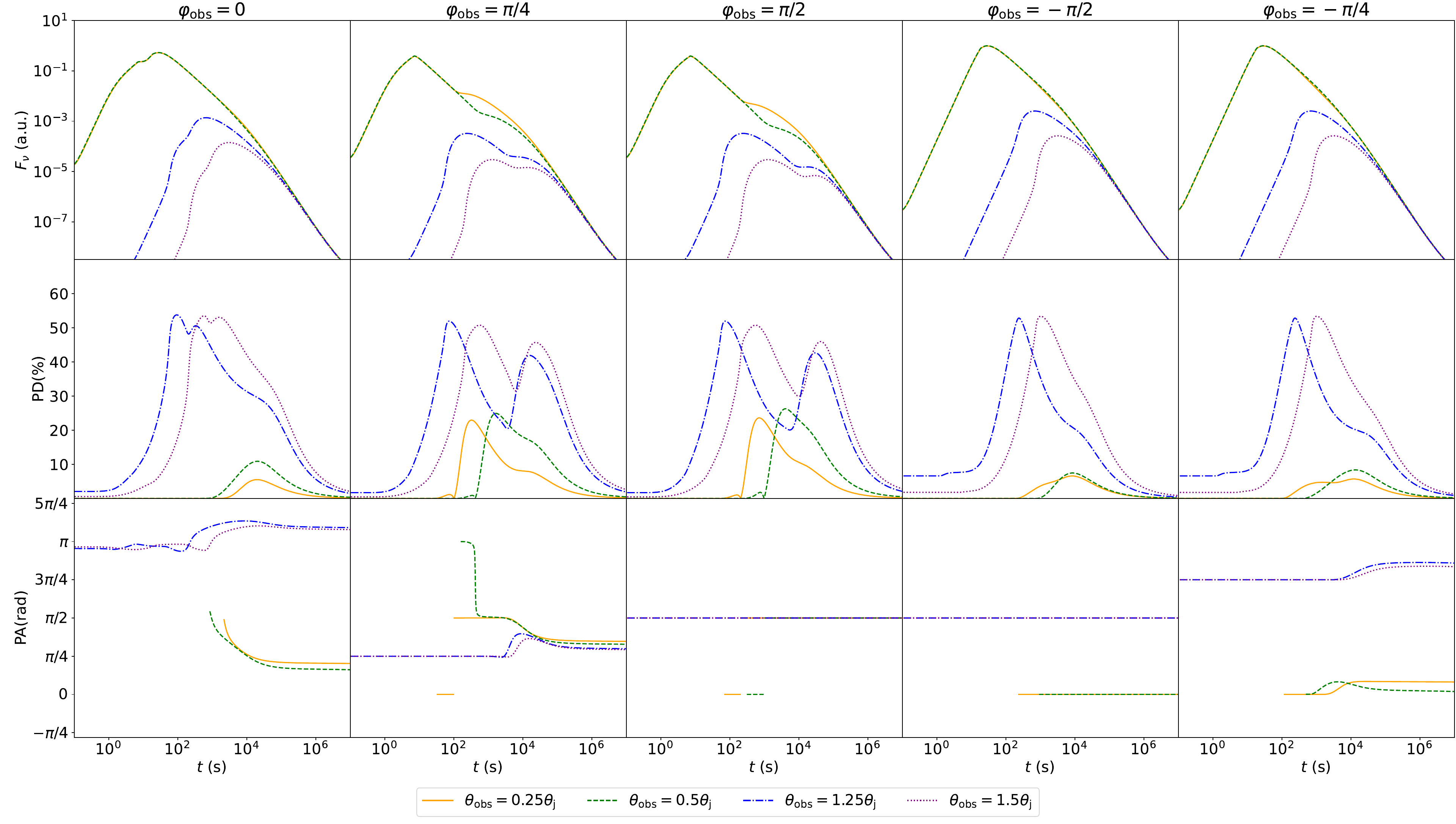}}
    \caption{The afterglow light curves ($F_{\nu}$) and the evolution of polarization degree (PD) and angle (PA) of the asymmetric jets with 2 elements at $\nu=8.22\times10^{14}$Hz. Polarization is only sensitive to the evolution of the light curve, so the flux of afterglow radiation has been normalized. Considering that low polarization degree is difficult to observe, we only show the polarization angle evolution when the polarization is significant. Generally, the range of polarization angle is $\left[0, \pi\right]$, but for the continuity of the polarization angle evolution curve, the range of the coordinate axis is extended. We set the Lorentz factors and the equivalent isotropic energy of the two elements as $\gamma_{0,1}=300$, $E_{\text{iso},1}=10^{51}$ergs and $\gamma_{0,2}=150$, $E_{\text{iso},2}=10^{52}$ergs, respectively. The relative position between the observer and the jet can be divided into three situations: (1)$\varphi=0$, the LOS through the plane of the interface; (2)$\varphi>0$ represent the LOS leans towards the first element; (3)$\varphi<0$ represent the LOS leans towards the second element. And different colors distinguish different $\theta_{\rm obs}$. Other parameters: $\theta_{\rm j}=5^{\circ}$, $\epsilon_e=0.01$, $\epsilon_B=0.001$, $p=2.7$, $n=10\text{cm}^{-3}$, $z=1$, $P_0=60\%$.}
    \label{2E0}
\end{figure*}

\begin{figure*}[htbp]
    \centering
    \subfigure{\includegraphics[width=17cm]{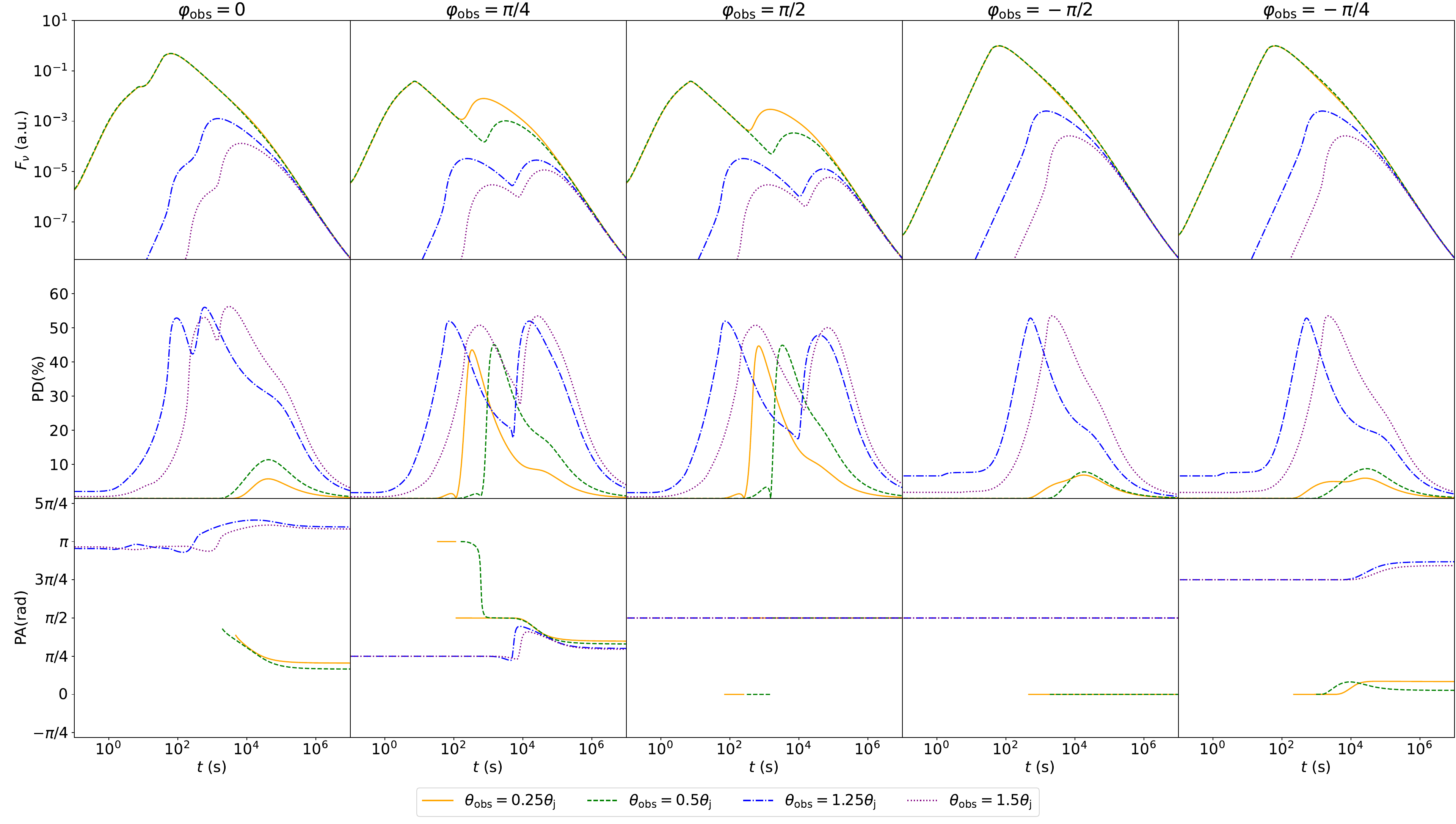}}
    \caption{Same to the Figure \ref{2E0}, but the $E_{\text{iso},2}=10^{53}$ergs.}
    \label{2Ee}
\end{figure*}

\begin{figure*}[htbp]
    \centering
    \subfigure{\includegraphics[width=17cm]{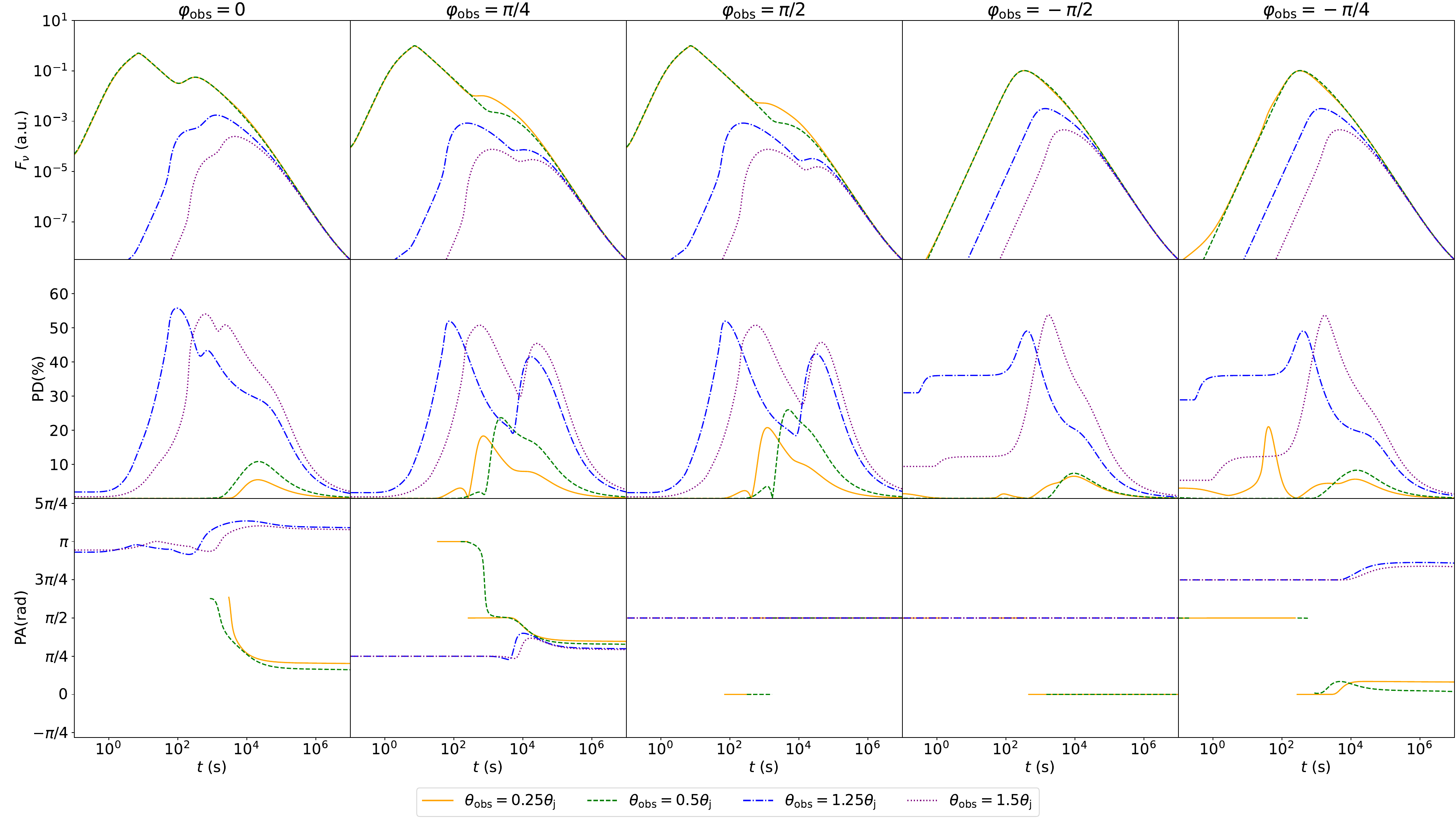}}
    \caption{Same to the Figure \ref{2E0}, but the $\gamma_{0,2}=60$.}
    \label{2Eg}
\end{figure*}

The afterglow light curves and the evolution of polarization degree and polarization angle of the asymmetric jets with 2 elements are shown in Figure \ref{2E0}, \ref{2Ee} and \ref{2Eg}. In the examples, we set $\gamma_{0,1}=300$, $\gamma_{0,2}=150$, $E_{\text{iso},1}=10^{51}$ergs, $E_{\text{iso},2}=10^{52}$ergs for Figure \ref{2E0}. In order to compare the asymmetric jets with different parameters, we set the $E_{\text{iso},2}=10^{53}$ergs for Figure \ref{2Ee} and $\gamma_{0,2}=60$ for Figure \ref{2Eg}. For each case, the relative position between the LOS and the jet can be divided into three situations: (1)$\varphi=0$, the LOS being through the plane of the interface; (2)$\varphi>0$, the LOS leaning towards the first element; (3)$\varphi<0$, the LOS leaning towards the second element. Specifically, we fix the half-opening angle $\theta_{\rm j}=5^\circ$. And we compare the polarization at different angle between LOS and jet axis $\theta_{\text{obs}}=0.25\theta_{\rm j}, 0.5\theta_{\rm j}, 1.25\theta_{\rm j}$ and $ 1.5\theta_{\rm j}$, respectively. We set other parameters related to the forward shock as: the fractions of the total shock-generated internal energy that goes into the random magnetic field $\epsilon_B=0.001$ and into the electrons $\epsilon_e=0.01$; particle number density of interstellar medium $n=10~\text{cm}^{-3}$; the power-law distribution index of electrons $p=2.7$; the red shift $z=1$ and the degree of polarization observed parallelly $P_0=60\%$ for synchrotron radiation. As an example, we consider an observation frequency of $\nu_{\rm obs}=8.22\times10^{14}$Hz for all the calculations of temporal evolution.

\begin{figure*}[htbp]
    \centering
    \subfigure{\includegraphics[width=17cm]{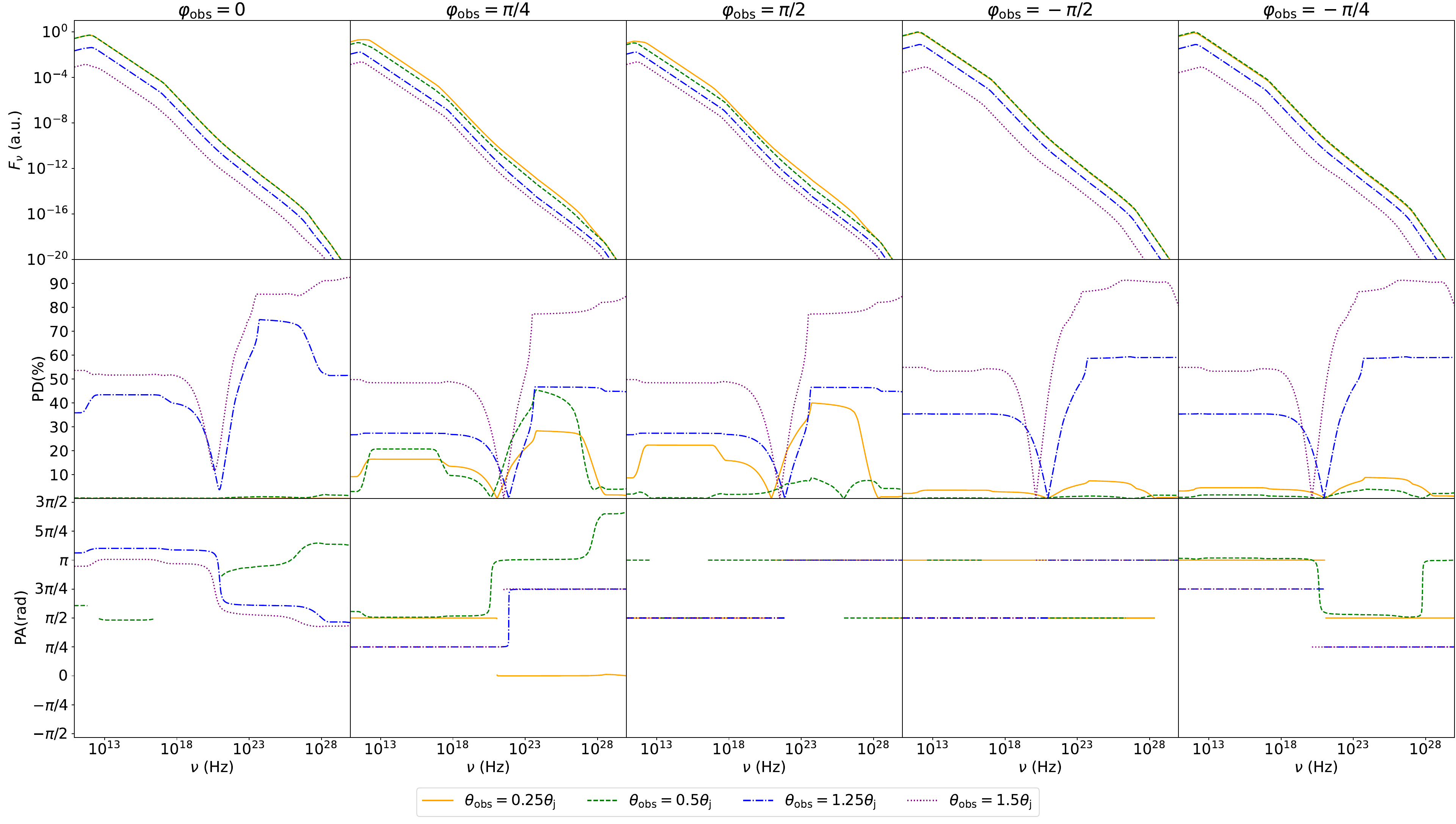}}
    \caption{The afterglow spectrum ($F_{\nu}$) and the spectral distribution of polarization degree (PD) and angle (PA) of the asymmetric jets with 2 elements at $t=10^3$s. Polarization is only sensitive to the shape of spectrum, so the flux of afterglow radiation has been normalized. The parameters are same to the Figure \ref{2E0}.}
    \label{2E0_spec}
\end{figure*}

\begin{figure*}[htbp]
    \centering
    \subfigure{\includegraphics[width=17cm]{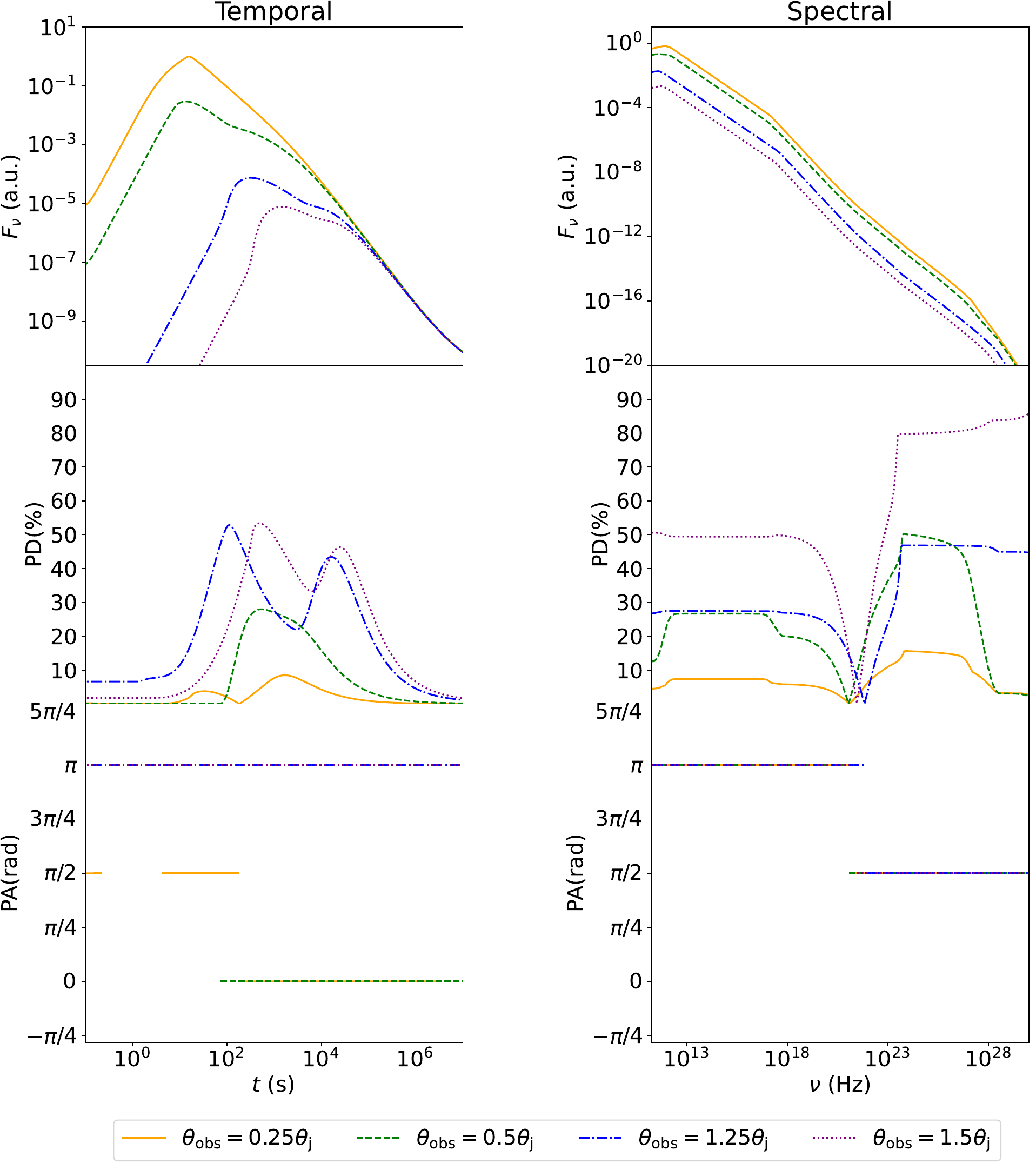}}
    \caption{As a comparison, we show the temporal evolution at $\nu=8.22\times10^{14}$Hz and spectral distribution at $t=10^3$s of afterglow and polarization of 2-component jets. We set the Lorentz factor $\gamma_{0,\text{inner}}=100$ and the equivalent isotropic energy $E_{\text{iso},\text{inner}}=10^{51}$ergs for inner component. And $\gamma_{0,\text{outer}}=50$, $E_{\text{iso},\text{outer}}=10^{50}$ergs for outer component. Other parameters are same to the Figure \ref{2E0}}
    \label{2C}
\end{figure*}

Our results suggest that for the asymmetric jets with 2 elements, polarization is generally present in the afterglow signal observed at most LOS, except for the direction along the jet axis. And the specific polarization evolution depends on the physical parameters of these two elements and the relative position between the jet and LOS. The element that dominate radiation can also dominate polarization evolution. 

As shown in Figure \ref{2E0}, overall, the polarization degree of the signal is significantly higher when the observation direction is off-axis compared to when the LOS is within the jet. For the latter case, the afterglow signal shows significant polarization at the late stage corresponding to the peak of the slower-moving component radiation, with a maximum polarization degree of up to $20\%$ when the LOS tends towards the faster-moving component. On the contrary, when the LOS tends towards the slower-moving component, the polarization may be observed with a maximum polarization degree of about $10\%$, which occurs later than when the LOS tends towards the faster-moving component. For the comparison of the cases with $\varphi_{\text{obs}}=\pm \pi/4$, when $\varphi_{\text{obs}}=\pi/4$, the maximum polarization degree corresponding to $\theta_{\text{obs}}=0.25\theta_{\text{j}}$ occurs at $t\sim320$s, while the maximum polarization degree corresponding to $\theta_{\text{obs}}=0.5\theta_{\text{j}}$ occurs at $t\sim1600$s, with both maximum polarization degrees being approximately $25\%$. Conversely, when $\varphi_{\text{obs}}=\pi/4$, regardless of whether $\theta_{\text{obs}}=0.25\theta_{\text{j}}$ or $\theta_{\text{obs}}=0.5\theta_{\text{j}}$, their maximum polarization degrees occur after $10^4$s and are both below $10\%$. In this case, when the LOS is perpendicular to the boundary between the faster- and slower-moving components (e.g. $\varphi_{\text{obs}}=\pm\pi/2$), the polarization angle does not change with time. Otherwise, the polarization angle will gradually rotate towards the direction of the component that is away from LOS after $1/\gamma_{1\text{ or }2}>\theta_{\rm obs}-\theta_{j}$. For example, when $\varphi_{\text{obs}}=\pi/4$, the polarization angle for $\theta_{\text{obs}}=0.25\theta_{\text{j}}$ or $0.5\theta_{\text{j}}$ rotates from approximately $90^{\circ}$ to about $60^{\circ}$ around $10^{4}$s towards the slower-moving component. In contrast, when $\varphi=-\pi/4$, the polarization angle rotates from approximately $0^{\circ}$ to about $15^{\circ}$ towards the faster-moving component before $10^{4}$s.

When the LOS falls outside the jet, polarization is present in the early rising phase of the light curve. When the LOS tends towards the slower-moving component, the early polarization can approach $10\%$, and as the main radiation of the jet gradually enters the LOS, the polarization evolution curve exhibits a peak corresponding to the light curve, with polarization reaching a maximum of $50\%$. In this case, if the LOS is perpendicular to the boundary between the faster- and slower-moving components, the polarization angle does not change with time. Otherwise, it will keep constant until the peak of the radiation from the slow-moving component, and then gradually rotate towards the direction of the faster-moving component. The polarization evolution for $\varphi_{\text{obs}}=-\pi/4$ and $\theta_{\text{obs}}=1.25\theta_{\text{j}}$ in Figure \ref{2E0} serves as a typical example. At $t\sim10$s, its polarization degree is about $8\%$, gradually rising to a maximum polarization degree of approximately $53\%$ by $t\sim240$s. The polarization angle remains about $135^{\circ}$ before $3000$s, after which it gradually rotates to approximately $155^{\circ}$ during the decay phase of the afterglow radiation. On the other hand, when the LOS tends towards the faster-moving component or is on the interface of two elements, the early polarization is below $5\%$, and as the main radiation of both faster- and slower-moving components enters the LOS successively, two peaks corresponding to the light curve appear on the polarization evolution curves, with the first higher peak reaching a maximum polarization of about $50\%$. Afterwards, the second lower peak reaching a maximum polarization of about $40\%$. In these cases, again the polarization angle does not change with time when $|\varphi_{\text{obs}}|=\pi/2$. Otherwise, it will almost keep constant until the peak of the radiation from the faster-moving component, and then gradually rotate towards the direction of the slower-moving component. However, after the slower-moving component dominants the radiation, the polarization angle rotates rapidly in the direction of faster-moving component, until as radiation decreases, the polarization angle gradually rotates again in the direction of the slower-moving component. Taking the polarization evolution for $\varphi_{\text{obs}}=\pi/4$ and $\theta_{\text{obs}}=1.25\theta_{\text{j}}$ as an example, its polarization degree is approximately $2\%$ at $1$s, reaching the first peak of $\sim52\%$ at about $71$s. Following this, the polarization degree gradually declines to a local minimum of about $20\%$, during which the polarization angle rotates from approximately $45^{\circ}$ to about $70^{\circ}$. The polarization degree reaches the second peak of approximately $42\%$ at $t\sim15700$s, while the polarization angle simultaneously rotates in the opposite direction to about $55^{\circ}$.

Next, we investigated the impact of increasing the energy ratio and velocity ratio of the two components on the results. As shown in Figure \ref{2Ee}, when the energy of the slower-moving component is increased, the re-brightening signature in the light curve becomes more prominent and the signal polarization during the rebrightening signal also increases, regardless of whether the LOS is within the jet or not. Consider the case of $\varphi_{\text{obs}}=0$ and $\theta_{\text{obs}}=1.25\theta_{\text{j}}$; in this scenario, the second peak of polarization degree in Figure \ref{2E0} is lower than the first, while in Figure \ref{2Ee}, it is the opposite. The evolution of polarization angle is not sensitive to the changes in energy ratio. As shown in Figure \ref{2Eg}, if the velocity ratio of the two components is further increased, the polarization evolution may become more complex. Especially when the LOS tends towards the slower-moving component, for the on-axis case, polarization signals may appear during the early rise phase of the light curve, with polarization degrees up to $20\%$, which is much higher than other on-axis cases at this phase. And the evolution at late-stage polarization are similar to the other cases. As shown in the case of $\varphi_{\text{obs}}=-\pi/4$ and $\theta_{\text{obs}} = 0.25\theta_{\text{j}}$, the maximum polarization degree occurs before the peak of the light curve. For the off-axis case, the polarization degree increases rapidly in the early stages and remains at a relatively high level (e.g. $20\%-40\%$) until the components enter the LOS. For instance, when $\varphi_{\text{obs}}=-\pi/4$ and $\theta_{\text{obs}} = 1.25\theta_{\text{j}}$, the polarization degree remains at $\sim40\%$ from $\sim1$s to $\sim100$s. The evolution of polarization angle is similar to the Figure \ref{2E0}.

Besides the temporal evolution, Figure \ref{2E0_spec} shows the spectral distribution of the afterglow's flux and polarization at $t=10^3$ s, utilizing the same set of parameters as in Figure \ref{2E0}.
%spectral distribution of afterglow and polarization at $t=10^3$s is shown in Figure \ref{2E0_spec}. 
%The parameters as same as in the Figure \ref{2E0}. 
Similar to the Top-Hat jet model, the polarization degree fluctuates at the breaks (characteristic frequencies) of the spectrum, such as $\nu_m$ and $\nu_c$. This phenomenon is observed in both synchrotron-dominated and SSC-dominated regimes. However, the fluctuation of polarization degree in the Top-Hat jet scenario is just a few percent, which is caused by the effect of equal-arrive-time surface. In contrast, under the combined influence of the equal-arrive-time surface and the jet's asymmetric structure, the polarization degree fluctuations of the 2-element jet can exceed $10\%$ in certain cases. Furthermore, as shown in Figure \ref{2E0_spec}, when the line of sight (LOS) lies within the jet, the extent of the polarization degree fluctuation is greater when the LOS is positioned at the faster-moving component ($\varphi_{\text{obs}} > 0$) compared to when it is located at the interface ($\varphi_{\text{obs}} = 0$) or the slower-moving component ($\varphi_{\text{obs}} < 0$). For example, consider the case with $\theta_{\text{obs}}=0.5\theta_{\text{j}}$. When $\varphi_{\text{obs}}=\pi/4$, the polarization degree initially increases from less than $10\%$ at $\nu_m$ (slightly greater than $10^{11}$Hz) to approximately $20\%$ at $10^{13}$Hz. As the frequency further increases, particularly in the vicinity of $\nu_c$ ($\sim10^{17}$Hz), the polarization degree decreases from $\sim20\%$ to $\sim10\%$.
%And around the $\nu_c$ ($\sim10^{17}$Hz), as the frequency increases, the polarization degree decreases from $\sim20\%$ to $\sim10\%$. 
In the frequency range where the polarization is SSC-dominated, having transitioned from being synchrotron-dominated, the polarization degree decreases from about $30\%$ at $10^{23}$Hz to only a few percent at $10^{27}$Hz. When $\varphi_{\text{obs}}=-\pi/4$, the polarization degree is close to $0\%$ across the entire spectrum, with the maximum fluctuation being less than $2\%$. As for the cases that the LOS are outside the jet ($\theta_{\rm obs} > \theta_j$), the shape of spectral distribution of polarization degree remains similar as those inside the jet.
%the extent of polarization degree fluctuations is similar.
Accordingly, the change in polarization degree at the characteristic frequencies, which could be accompanied by the rotation of polarization angle, may serve as an important evidence of the existence of a 2-element jet. For example, when $\varphi_{\text{obs}}=\pi/4$ and $\theta_{\text{obs}}=0.5\theta_{\text{j}}$, the polarization angle rotated by approximately $10^{\circ}$ around $\nu_m$. 
%, otherwise the polarization angle would remain unchanged.

When the observation frequency exceeds approximately $10^{18}$ Hz, SSC scattering becomes non-negligible. Due to the disparity in $\theta_p$ between synchrotron and SSC emissions \citep{Gill2020MNRAS}, the polarization degree decreases as the frequency increases, reaching a local minimum at approximately $\sim10^{21}$ Hz, which accompanied by an intense rotation of the polarization angle by approximately $\pi/2$.
%On either side of the frequency at which the polarization degree reaches the local minimum, the polarization angles differ by approximately $\sim\pi/2$. 
Moreover, for all structures of jets, when the observation frequency is higher than the frequency at which the degree of polarization reaches its local minimum, the radiation from SSC dominates the polarization. Since $P_0$ in SSC scenario is almost $100\%$ in all cases \citep{Gill2020MNRAS}, its degree of polarization usually achieves a higher level than that of synchrotron radiation. Notably, for the 2-element jet, in most cases, the local minimum of the degree of polarization is $\sim0\%$, and the polarization angle undergoes a sudden change, similar to that with a Top-Hat jet. However, in certain situations, the local minimum of the polarization degree is $>0\%$, and the change in the polarization angle is intense but continuous. For example, when $\varphi=0$ and $\theta_{\text{obs}}=1.5\theta_{\text{j}}$, the local minimum of the polarization degree is approximately $10\%$. This phenomenon can be attributed to the influence of the radiation from two asymmetric elements on the symmetry of the radiation area when the LOS approaches the boundary between the two elements.

In previous works, another axisymmetric two-component jet, composed of an inner and an outer part, has often been discussed \citep[e.g.][]{Wu2005MNRAS}. The inner part consists of a jet with a small opening angle and high velocity, while the outer part consists of a jet with a large opening angle and low velocity. It is of great interest to compare the emission and polarization properties of this jet structure with the non-axisymmetric two-component jet discussed here. We constructed an axisymmetric two-component jet with the following parameters: half-opening angle of outer jet is $\theta_{\text{j}}=5^\circ$; the half-opening angle of the inner component is $\theta_{\text{c}}=0.3\theta_{\text{j}}$; the Lorentz factor and equivalent isotropic kinetic energy of the inner component are $\gamma_{0,\text{inner}}=300$ and $E_{\text{iso,inner}}=10^{52}$ergs, respectively; the Lorentz factor and equivalent isotropic kinetic energy of the outer component are $\gamma_{0,\text{outer}}=150$ and $E_{\text{iso,outer}}=10^{51}$ergs. As shown in Figure \ref{2C}, such a two-component jet also produces rebrightening signature in the light curve. When the LOS is within the jet, the afterglow signal will show significant polarization at the late stage when $1/\gamma_{\text{inner or outer}}>\text{min}(|\theta_{\text{c}}-\theta_{\text{obs}}|, \theta_{\text{j}}-\theta_{\text{obs}})$, with a maximum polarization degree of up to $30\%$. When the LOS is outside the jet, polarization is present in the early rising phase of the light curve, and the early polarization can be greater than $0\%$. As the radiation of the outer and inner jet entering the LOS one after another, the polarization evolution curve exhibits two peak corresponding to the light curve, with the polarization reaching a maximum of $50\%$. In the scenario of axisymmetric two-component jet, the azimuth angle of the LOS direction does not affect the magnitude of the polarization degree, but only determines the specific direction of the polarization angle. Compared to Figure \ref{2E0}, Figure \ref{2E0_spec} and Figure \ref{2C}, we find that there is a certain degree of similarity between the light curve and polarization evolution curves of axisymmetric and non-axisymmetric two-component jets. Their spectral distributions of flux and polarization degree are also almost indistinguishable in most cases, except at the frequency where both the contributions of synchrotron and SSC radiations to the polarization are non-negligible. Detecting a local minimum polarization degree that is greater than zero in such cases can provide evidence for the existence of asymmetric structures. In addition, to effectively distinguish between these two structures, monitoring the temporal evolution or spectral distribution of the polarization angle is necessary. For axisymmetric jets, the polarization angle remains in two directions: perpendicular or parallel to the plane of the jet axis and the LOS, while for non-axisymmetric jets, the polarization angle may exhibit complex changes. 

\subsection{More than 2 elements in a jet}

For more complex asymmetric structures, the jet may be divided into multiple elements with $N>2$. These individual elements exhibit differences in physical parameters, which may consequently lead to more complex polarization evolution and spectral distribution. 

\begin{figure}[htbp]
	\centering
	\subfigure{\includegraphics[width=8.5cm]{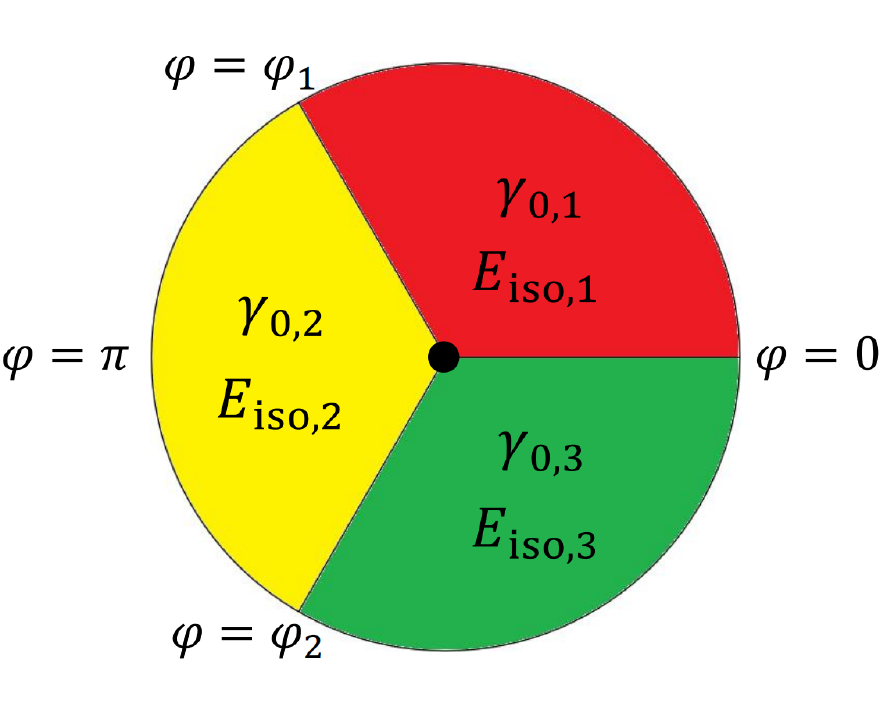}
		\label{3.2}}
	\caption{The schematic diagrams of the cross-section of a asymmetric jet with 3 elements. $\varphi=0, \varphi_1, \varphi_2$ are the interfaces of the jet with 3 elements.}
	\label{3.23}
\end{figure}

\begin{figure*}[htbp]
	\centering
	\includegraphics[width=17cm]{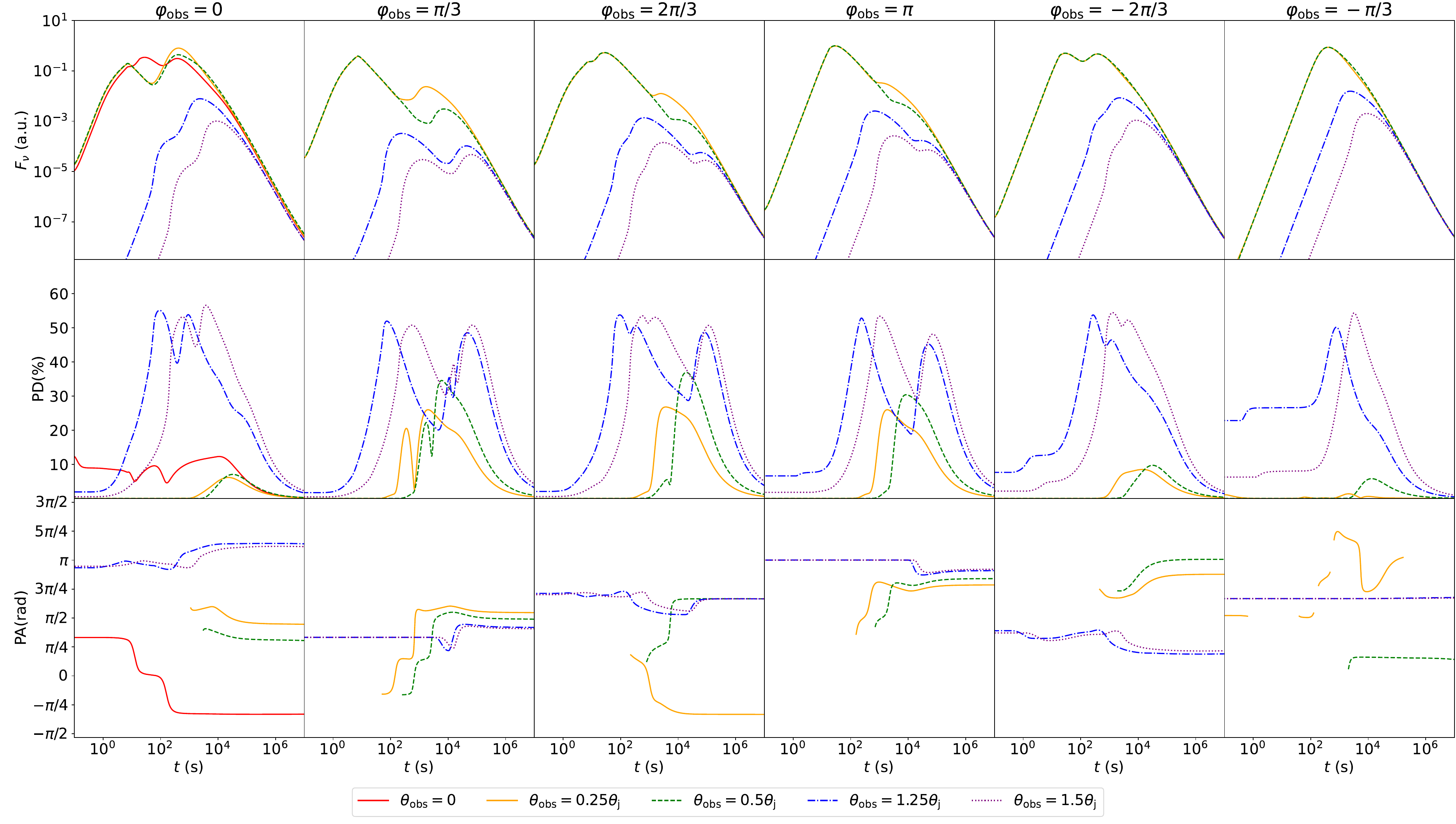}
	
        \caption{The afterglow light curves ($F_{\nu}$) and the evolution of polarization degree (PD) and angle (PA) of the asymmetric jets with 3 elements at $\nu=8.22\times10^{14}$Hz. We show the light curves and polarization evolution in the range from $\varphi_{\text{obs}}=0$ to $\varphi_{\text{obs}}=5\pi/3$, and from $\theta_{\text{obs}}=0$ to $\theta_{\text{obs}}=1.5\theta_{\text{j}}$. The flux of afterglow radiation has been normalized. The interfaces of the 3 elements are at $\varphi=0$, $2\pi/3$ and $4\pi/3$. The initial Lorentz factor and equivalent isotropic kinetic energy of each element are $\gamma_{0,1}=100$, $E_{\text{iso},1}=10^{50}$ergs; $\gamma_{0,2}=50$, $E_{\text{iso},2}=10^{51}$ergs; $\gamma_{0,3}=25$, $E_{\text{iso},3}=10^{52}$ergs. Other parameters are same to the Figure \ref{2E0}.}
        \label{3E3}
\end{figure*}

As an example, Figure \ref{3.23} shows the schematic diagrams of the cross-sections of a asymmetric jets with 3 elements. 
And based on the different initial Lorentz factors $\gamma_0$ and equivalent isotropic kinetic energy $E_{\text{iso}}$ of each element, the mathematical expression for their structures are
\begin{equation}
    \gamma_0=\begin{cases}
    \gamma_{0,1}&0<\varphi<\varphi_1,\\
    \gamma_{0,2}&\varphi_1<\varphi<\varphi_2,\\
    \gamma_{0,3}&\text{others},
    \end{cases}
    \label{g03}
\end{equation}
\begin{equation}
    E_{\text{iso}}=\begin{cases}
    E_{\text{iso},1}&0<\varphi<\varphi_1,\\
    E_{\text{iso},2}&\varphi_1<\varphi<\varphi_2,\\
    E_{\text{iso},3}&\text{others},
    \end{cases}
    \label{E03}
\end{equation}
for the jet with 3 elements. And the evolution of their corresponding afterglow light curves, polarization degree and polarization angle at $\nu=8.22\times10^{14}$Hz are shown in Figure \ref{3E3}. We set $\gamma_{0,1}=300$, $E_{\text{iso},1}=10^{51}$ergs; $\gamma_{0,2}=150$, $E_{\text{iso},2}=10^{52}$ergs; $\gamma_{0,3}=75$, $E_{\text{iso},3}=10^{53}$ergs for the jet with 3 elements. Other parameters are same to the Figure \ref{2E0}. We considered both the on-axis and off-axis observation, and $\varphi_{\text{obs}}$ ranges from $-\pi$ to $\pi$. 

The results indicate that for asymmetric jets with $N>2$ elements, polarization can be observed from any direction. As shown in Figure \ref{3E3}, when the LOS is along the jet axis, the afterglow signal shows a polarized signal from the early stages, and the polarization evolution curve corresponds to fluctuations in the light curve. The polarization degree can reach up to $10\%$. When the LOS is pointing towards a different direction within the jet, the afterglow signal will show polarization at a later time. The polarization degree evolution curve will also fluctuate corresponding to the evolution of the light curve. The maximum polarization degree can reach $20\%-50\%$, which is essentially related to the specific parameters (especially the velocity) of the jet element the line of sight falling in. 

When the LOS is outside the jet, polarization is present in the early rising phase of the light curve. For the cases whose LOS tends towards the faster-moving components, the early polarization degree is relatively small ($<10\%$), and there are multiple peaks in the polarization degree evolution curves, with the maximum polarization degree reaching up to $50\%$. For the cases whose LOS tends towards the slower-moving components, the polarization degree at early time could significantly increase, even up to $30\%$. In these cases, after maintaining a high level of polarization for a period of time, there may be fluctuations or even higher peaks in the late stage, or it may gradually decay depending on the comparison of different elements' parameters. 

For both on-axis and off-axis scenarios, the polarization angle will undergo significant deviation as the radiation intensity of different elements alternate dominance. The speed of polarization angle rotation depends on the difference in parameters between adjacent elements, which essentially reflects the degree of turbulence in the jet structure. 

\begin{figure*}[htbp]
	\centering
	\includegraphics[width=17cm]{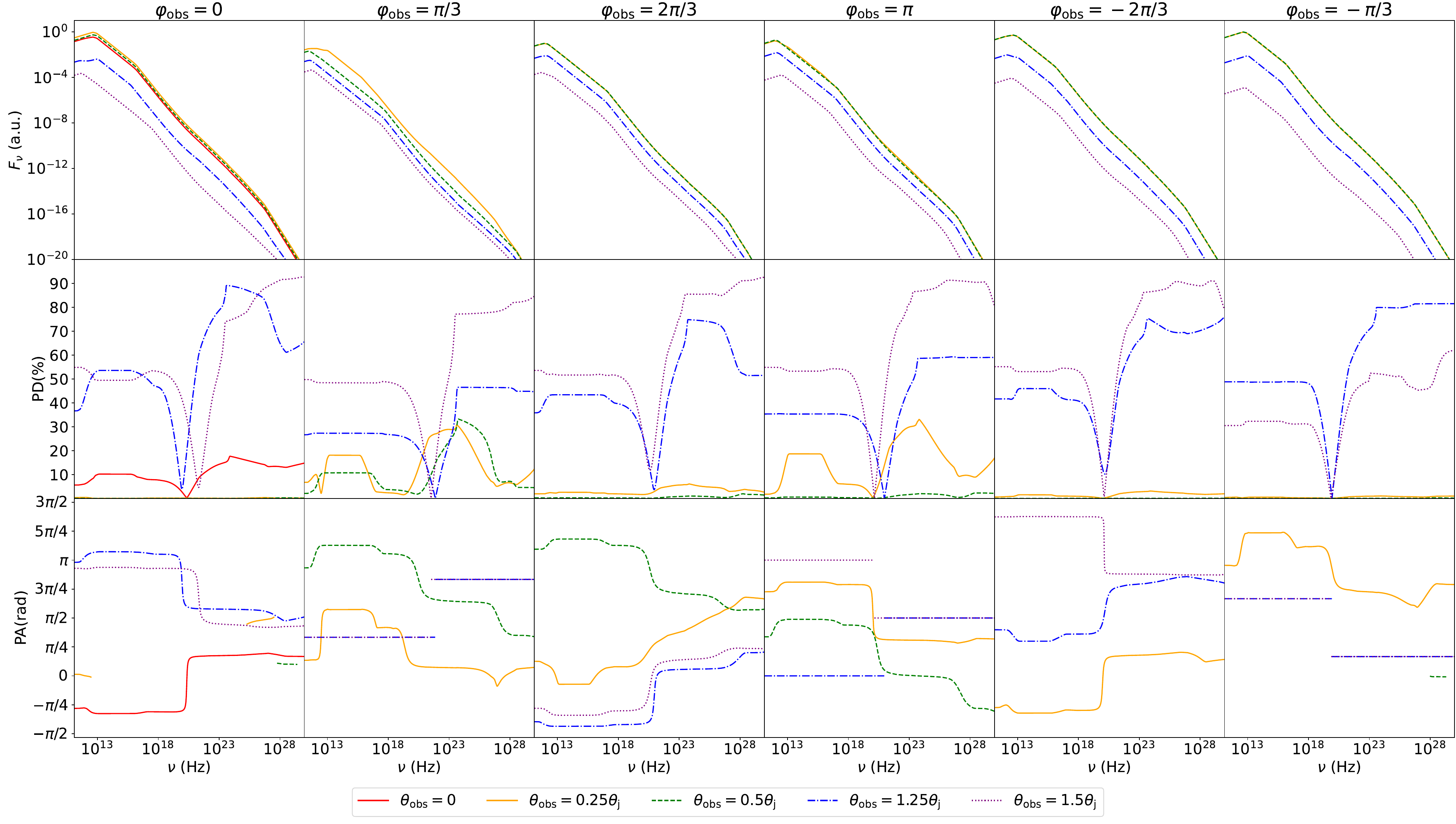}
	
        \caption{The spectrum ($F_{\nu}$) and the spectral distribution of polarization degree (PD) and angle (PA) of the asymmetric jets with 3 elements at $t=10^3$s. Other parameters are same to the Figure \ref{3E3}.}
        \label{3E_spec}
\end{figure*}

Figure \ref{3E_spec} shows the spectral distribution of the afterglow's flux and polarization at $t=10^3$ s, for the 3-element jet, utilizing the same set of parameters as in Figure \ref{3E3}. Compare to Figure \ref{2E0_spec}, the rotation of the polarization angle is more pronounced around the characteristic frequency. However, the spectral distributions of both flux and polarization are generally similar to those of the 2-element jet. In rare instances can they be distinctly differentiated from the 2-element jet. For example, when $\varphi_{\text{obs}}=2\pi/3$ and $\theta_{\text{obs}}=0.25\theta_{\text{j}}$, in the SSC-dominated bands, the polarization angle shows a persistent variation across the entire frequency range. However, in many cases, such as when $\varphi_{\text{obs}}=\pi/3$ and $\theta_{\text{obs}}=0.5\theta_{\text{j}}$, the spectral distribution characteristics of the 2-element jet still apply to the 3-element jet. Therefore, the spectral distribution of complex asymmetric structures is often difficult to distinguish from that of 2-element jets. This indicates that a more complex asymmetric structure does not necessarily make the spectral distribution of flux and polarization more complex. This is mainly because, in most cases, the afterglow is dominated by a certain element. Radiation from other elements cannot significantly affect the afterglow and polarization. Therefore, to distinguish between different asymmetric structures, it is necessary to monitor the temporal evolution of afterglow and polarization.

In summary, when compared to the situation with 2-elements, there are several points worthy of note when the jet structure is more complex: 1) for the 2-element case, if the LOS is exactly along the jet axis, the polarization degree would be zero. But for multi-element cases, a polarization signal will still appear even $\theta_{\rm obs}=0$; 2) the evolution curve of polarization degree and polarization angle will have more fluctuations together with the light curve; 3) the maximum polarization degree, which is around 50\%, does not significantly increase due to the complexity of the jet structure. Moreover, due to the similarity in the spectral distribution of flux and polarization among different asymmetric structures, monitoring their temporal evolution is necessary to distinguish between these structures.

\section{Conclusion and discussion}

In this work, we analyzed the polarization characteristics of gamma-ray burst afterglows in the context of a non-axisymmetric structured jet, where the jet is divided into $N$ independent and uniform "patch" based on the changes in physical parameters along the azimuth angle ($\varphi$) direction. Our results reveal a significant impact of asymmetric structures on polarization evolution and spectral distribution.

The evolution of polarization produced by a non-axisymmetric jet is related to the specific structure of the jet and the direction of the LOS. The polarization degree is generally higher when the LOS is outside the jet than when it is inside the jet. According to the evolution, similar to the characteristics of light curves, the polarization degree also experiences fluctuations as different regions of radiation alternate in dominance, accompanied by gradual rotation of the polarization angle. Relatively low polarization degree and rapid rotation of the polarization angle usually occurs when the contributions of radiation from different regions are competing. Therefore the complexity of the polarization evolution reflects the complexity of the jet structure. Besides the temporal evolution, the spectral distribution of polarization also emphasizes the influence of the jet structure. For the afterglow produced by a non-axisymmetric jet, the fluctuation in polarization degree at the breaks of the spectrum is greater compared to that of a Top-Hat jet, and it could be accompanied by a rotation of the polarization angle. Additionally, both synchrotron and SSC radiation exhibit significant polarization. However, when the contributions from synchrotron and SSC radiation compete, the polarization degree can be reduced, resulting in a local minimum. The local minimum of the polarization degree generated by symmetric structures is always 0, whereas for asymmetric structures, the local minimum may be non-negligible.

Comparing different asymmetric structures, the polarization and radiation spectral distributions are generally similar. With few exceptions, it is difficult to distinguish different structures based solely on their spectral distribution. However, more complex asymmetric structures exhibit more intricate temporal evolution. Therefore, monitoring the evolution of light curves and polarization curves is more effective for distinguishing between asymmetric jets of varying complexity.

For the spectral distribution of polarization, we did not consider the frequency range below the synchrotron self-absorption frequency ($\nu_a$), where a significant fraction of electrons become thermally distributed. Therefore, the results in the frequency range below $\nu_a$ should be modified accordingly \citep{Warren2018MNRAS}.
%present the afterglow emission below the synchrotron self-absorption frequency ($\nu_a$). 
%If the synchrotron self-absorption process is significant, thermal electrons have a substantial impact.  
\citet{Mao2018ApJ} have analyzed the impact of relativistic thermal electrons on synchrotron polarization. They found that the large optical depth for radiative transfer results in a low degree of polarization. Future works incorporating the synchrotron self-absorption effect into the framework with a non-axisymmetric jet can aid in studying the jet structure using polarization observations from lower frequencies.
%Since the primary goal of this paper is to identify the polarization characteristics of asymmetric structures, providing precise predictions of the effects of synchrotron self-absorption and thermal electron populations on the polarization of GRB afterglows is beyond the scope of this work.

It is worth noting that the specific value of the polarization degree shown in this work is related to the parameter selection used in the examples provided. In real observations, the afterglow signal of different GRBs may exhibit different polarization degrees. However, the evolution of polarization degree and polarization angle shown in this work's results should be universally applicable. 

Although it is difficult to observe the polarization of GRBs, many valuable polarization observations have still been achieved so far. For example, the polarization degree of $10.9\%$, $28\%$ and $<6\%$ were observed during the early afterglow phase of GRB 090102 \citep{Steele2009Nature}, GRB 120308A \citep{Mundell2013Nature} and GRB 190829A \citep{Dichiara2022MNRAS}, respectively. For the late afterglow phase, the polarization degree of $<7\%$, $0.27\%$, $0.8\%$ and $1.2\%$ were observed in GRB 991216 \citep{Granot2005ApJ}, GRB 171205A \citep{Urata2019ApJ}, GRB 190114C \citep{Laskar2019ApJ} and GRB 191221B \citep{Buckley2021MNRAS,Urata2023NatAs}, respectively. In recent years, with the development of polarization related telescopes and observation technologies, such as POLAR project \citep{Orsi2011ASTRA,ORSI2011ICRC}, MOPTOP \citep{Shrestha2020MNRAS}, and MASTER \citep{Lipunov2019ARep}, the polarization of gamma-ray bursts has garnered more attention from researchers. In the future, increased polarization observations of GRB prompt emissions and afterglows will shed light on the mechanisms responsible for their polarization.

%Based on observations of the afterglow light curve, it is found that some GRBs do have fluctuations or multiple rebrightening signatures during the late stages \citep{deWet2023A&A}. Possible explanations for these phenomena include late-stage energy injection or asymmetric structured jets as mentioned here. Since the energy injection process does not significantly affect polarization, the results of this work suggest that polarization detection can effectively distinguish between the two models. On the other hand, if there is only one significant rebrightening behavior in the late afterglow and it corresponds to a significant polarization degree, it may also be caused by an axisymmetric structured jet \citep[e.g., the two-component jets][]{Huang2004ApJ,Peng2005ApJ,Wu2005MNRAS,Beniamini2020MNRAS}. In this case, the results of this work suggest that the symmetry of the jet structure can be determined by monitoring whether there is a rotation in the polarization angle. Overall, in the future, it is recommended to perform polarimetric observations on GRBs with special features in their late afterglows, in order to ultimately determine the symmetry of the GRB jet structure.

Based on observations of the afterglow light curve, it is found that some GRBs do have fluctuations or multiple rebrightening signatures during the late stages \citep{deWet2023A&A}. Possible explanations for these phenomena include late-stage energy injection, reverse shock or asymmetric structured jets as mentioned here. Since the energy injection process does not significantly affect polarization, the results of this work suggest that polarization detection can effectively distinguish between the two models. And regarding the reverse shock, \citet{Lan2016ApJL} have discussed the polarization in scenarios where either the forward shock or the reverse shock dominates. The ordered magnetic field in the reverse shock is markedly different from the tangled magnetic field in the forward shock. Generally, the reverse shock is distinguished by a sudden change in polarization angle at vicinity of the crossing time and after the jet break time. On the other hand, if there is only one significant rebrightening behavior in the late afterglow and it corresponds to a significant polarization degree, it may also be caused by an axisymmetric structured jet \citep[e.g., the two-component jets][]{Huang2004ApJ,Peng2005ApJ,Wu2005MNRAS,Beniamini2020MNRAS}. In this case, the results of this work suggest that the symmetry of the jet structure can be determined by monitoring whether there is a rotation in the polarization angle. Overall, in the future, it is recommended to perform polarimetric observations on GRBs with special features in their late afterglows, in order to ultimately determine the symmetry of the GRB jet structure.

\section*{Acknowledgements}

We appreciate the insightful comments and valuable suggestions from the anonymous referee. This work is supported by the National Natural Science Foundation of China (Projects 12373040,12021003, U2038107), the National SKA Program of China (2022SKA0130100), the China Postdoctoral Science Foundation (Project 2023M732713), the Fundamental Research Funds for the Central Universities and the Carlsberg Foundation (CF18-0183, PI: I. Tamborra). 

\bibliography{sample631}{}

\begin{thebibliography}{}
\expandafter\ifx\csname natexlab\endcsname\relax\def\natexlab#1{#1}\fi
\providecommand{\url}[1]{\href{#1}{#1}}
\providecommand{\dodoi}[1]{doi:~\href{http://doi.org/#1}{\nolinkurl{#1}}}
\providecommand{\doeprint}[1]{\href{http://ascl.net/#1}{\nolinkurl{http://ascl.net/#1}}}
\providecommand{\doarXiv}[1]{\href{https://arxiv.org/abs/#1}{\nolinkurl{https://arxiv.org/abs/#1}}}

\bibitem[{Abbott {et~al.}(2017)Abbott, Abbott, Abbott, Acernese, Ackley, Adams, Adams, Addesso, Adhikari, Adya, Affeldt, Afrough, Agarwal, Agathos, Agatsuma, Aggarwal, Aguiar, Aiello, Ain, Ajith, Allen, Allen, Allocca, Altin, Amato, Ananyeva, Anderson, Anderson, Angelova, Antier, Appert, Arai, Araya, Areeda, Arnaud, Arun, Ascenzi, Ashton, Ast, Aston, Astone, Atallah, Aufmuth, Aulbert, AultONeal, Austin, Avila-Alvarez, Babak, Bacon, Bader, Bae, Bailes, Baker, Baldaccini, Ballardin, Ballmer, Banagiri, Barayoga, Barclay, Barish, Barker, Barkett, Barone, Barr, Barsotti, Barsuglia, Barta, Barthelmy, Bartlett, Bartos, Bassiri, Basti, Batch, Bawaj, Bayley, Bazzan, B\'ecsy, Beer, Bejger, Belahcene, Bell, Berger, Bergmann, Bernuzzi, Bero, Berry, Bersanetti, Bertolini, Betzwieser, Bhagwat, Bhandare, Bilenko, Billingsley, Billman, Birch, Birney, Birnholtz, Biscans, Biscoveanu, Bisht, Bitossi, Biwer, Bizouard, Blackburn, Blackman, Blair, Blair, Blair, Bloemen, Bock, Bode, Boer, Bogaert, Bohe, Bondu, Bonilla, Bonnand,
  Boom, Bork, Boschi, Bose, Bossie, Bouffanais, Bozzi, Bradaschia, Brady, Branchesi, Brau, Briant, Brillet, Brinkmann, Brisson, Brockill, Broida, Brooks, Brown, Brown, Brunett, Buchanan, Buikema, Bulik, Bulten, Buonanno, Buskulic, Buy, Byer, Cabero, Cadonati, Cagnoli, Cahillane, Calder\'on~Bustillo, Callister, Calloni, Camp, Canepa, Canizares, Cannon, Cao, Cao, Capano, Capocasa, Carbognani, Caride, Carney, Carullo, Casanueva~Diaz, Casentini, Caudill, Cavagli\`a, Cavalier, Cavalieri, Cella, Cepeda, Cerd\'a-Dur\'an, Cerretani, Cesarini, Chamberlin, Chan, Chao, Charlton, Chase, Chassande-Mottin, Chatterjee, Chatziioannou, Cheeseboro, Chen, Chen, Chen, Cheng, Chia, Chincarini, Chiummo, Chmiel, Cho, Cho, Chow, Christensen, Chu, Chua, Chua, Chung, Chung, Ciani, Ciolfi, Cirelli, Cirone, Clara, Clark, Clearwater, Cleva, Cocchieri, Coccia, Cohadon, Cohen, Colla, Collette, Cominsky, Constancio, Conti, Cooper, Corban, Corbitt, Cordero-Carri\'on, Corley, Cornish, Corsi, Cortese, Costa, Coughlin, Coughlin, Coulon,
  Countryman, Couvares, Covas, Cowan, Coward, Cowart, Coyne, Coyne, Creighton, Creighton, Cripe, Crowder, Cullen, Cumming, Cunningham, Cuoco, Dal~Canton, D\'alya, Danilishin, D'Antonio, Danzmann, Dasgupta, Da~Silva~Costa, Dattilo, Dave, Davier, Davis, Daw, Day, De, DeBra, Degallaix, De~Laurentis, Del\'eglise, Del~Pozzo, Demos, Denker, Dent, De~Pietri, Dergachev, De~Rosa, DeRosa, De~Rossi, DeSalvo, de~Varona, Devenson, Dhurandhar, D\'{\i}az, Dietrich, Di~Fiore, Di~Giovanni, Di~Girolamo, Di~Lieto, Di~Pace, Di~Palma, Di~Renzo, Doctor, Dolique, Donovan, Dooley, Doravari, Dorrington, Douglas, Dovale~\'Alvarez, Downes, Drago, Dreissigacker, Driggers, Du, Ducrot, Dudi, Dupej, Dwyer, Edo, Edwards, Effler, Eggenstein, Ehrens, Eichholz, Eikenberry, Eisenstein, Essick, Estevez, Etienne, Etzel, Evans, Evans, Factourovich, Fafone, Fair, Fairhurst, Fan, Farinon, Farr, Farr, Fauchon-Jones, Favata, Fays, Fee, Fehrmann, Feicht, Fejer, Fernandez-Galiana, Ferrante, Ferreira, Ferrini, Fidecaro, Finstad, Fiori, Fiorucci,
  Fishbach, Fisher, Fitz-Axen, Flaminio, Fletcher, Fong, Font, Forsyth, Forsyth, Fournier, Frasca, Frasconi, Frei, Freise, Frey, Frey, Fries, Fritschel, Frolov, Fulda, Fyffe, Gabbard, Gadre, Gaebel, Gair, Gammaitoni, Ganija, Gaonkar, Garcia-Quiros, Garufi, Gateley, Gaudio, Gaur, Gayathri, Gehrels, Gemme, Genin, Gennai, George, George, Gergely, Germain, Ghonge, Ghosh, Ghosh, Ghosh, Giaime, Giardina, Giazotto, Gill, Glover, Goetz, Goetz, Gomes, Goncharov, Gonz\'alez, Gonzalez~Castro, Gopakumar, Gorodetsky, Gossan, Gosselin, Gouaty, Grado, Graef, Granata, Grant, Gras, Gray, Greco, Green, Gretarsson, Groot, Grote, Grunewald, Gruning, Guidi, Guo, Gupta, Gupta, Gushwa, Gustafson, Gustafson, Halim, Hall, Hall, Hamilton, Hammond, Haney, Hanke, Hanks, Hanna, Hannam, Hannuksela, Hanson, Hardwick, Harms, Harry, Harry, Hart, Haster, Haughian, Healy, Heidmann, Heintze, Heitmann, Hello, Hemming, Hendry, Heng, Hennig, Heptonstall, Heurs, Hild, Hinderer, Ho, Hoak, Hofman, Holt, Holz, Hopkins, Horst, Hough, Houston, Howell,
  Hreibi, Hu, Huerta, Huet, Hughey, Husa, Huttner, Huynh-Dinh, Indik, Inta, Intini, Isa, Isac, Isi, Iyer, Izumi, Jacqmin, Jani, Jaranowski, Jawahar, Jim\'enez-Forteza, Johnson, Johnson-McDaniel, Jones, Jones, Jonker, Ju, Junker, Kalaghatgi, Kalogera, Kamai, Kandhasamy, Kang, Kanner, Kapadia, Karki, Karvinen, Kasprzack, Kastaun, Katolik, Katsavounidis, Katzman, Kaufer, Kawabe, K\'ef\'elian, Keitel, Kemball, Kennedy, Kent, Key, Khalili, Khan, Khan, Khan, Khazanov, Kijbunchoo, Kim, Kim, Kim, Kim, Kim, Kim, Kimbrell, King, King, Kinley-Hanlon, Kirchhoff, Kissel, Kleybolte, Klimenko, Knowles, Koch, Koehlenbeck, Koley, Kondrashov, Kontos, Korobko, Korth, Kowalska, Kozak, Kr\"amer, Kringel, Krishnan, Kr\'olak, Kuehn, Kumar, Kumar, Kumar, Kuo, Kutynia, Kwang, Lackey, Lai, Landry, Lang, Lange, Lantz, Lanza, Larson, Lartaux-Vollard, Lasky, Laxen, Lazzarini, Lazzaro, Leaci, Leavey, Lee, Lee, Lee, Lee, Lee, Lehmann, Lenon, Leon, Leonardi, Leroy, Letendre, Levin, Li, Linker, Littenberg, Liu, Liu, Lo, Lockerbie, London,
  Lord, Lorenzini, Loriette, Lormand, Losurdo, Lough, Lousto, Lovelace, L\"uck, Lumaca, Lundgren, Lynch, Ma, Macas, Macfoy, Machenschalk, MacInnis, Macleod, Maga\~na Hernandez, Maga\~na Sandoval, Maga\~na Zertuche, Magee, Majorana, Maksimovic, Man, Mandic, Mangano, Mansell, Manske, Mantovani, Marchesoni, Marion, M\'arka, M\'arka, Markakis, Markosyan, Markowitz, Maros, Marquina, Marsh, Martelli, Martellini, Martin, Martin, Martynov, Marx, Mason, Massera, Masserot, Massinger, Masso-Reid, Mastrogiovanni, Matas, Matichard, Matone, Mavalvala, Mazumder, McCarthy, McClelland, McCormick, McCuller, McGuire, McIntyre, McIver, McManus, McNeill, McRae, McWilliams, Meacher, Meadors, Mehmet, Meidam, Mejuto-Villa, Melatos, Mendell, Mercer, Merilh, Merzougui, Meshkov, Messenger, Messick, Metzdorff, Meyers, Miao, Michel, Middleton, Mikhailov, Milano, Miller, Miller, Miller, Millhouse, Milovich-Goff, Minazzoli, Minenkov, Ming, Mishra, Mitra, Mitrofanov, Mitselmakher, Mittleman, Moffa, Moggi, Mogushi, Mohan, Mohapatra, Molina,
  Montani, Moore, Moraru, Moreno, Morisaki, Morriss, Mours, Mow-Lowry, Mueller, Muir, Mukherjee, Mukherjee, Mukherjee, Mukund, Mullavey, Munch, Mu\~niz, Muratore, Murray, Nagar, Napier, Nardecchia, Naticchioni, Nayak, Neilson, Nelemans, Nelson, Nery, Neunzert, Nevin, Newport, Newton, Ng, Nguyen, Nguyen, Nichols, Nielsen, Nissanke, Nitz, Noack, Nocera, Nolting, North, Nuttall, Oberling, O'Dea, Ogin, Oh, Oh, Ohme, Okada, Oliver, Oppermann, Oram, O'Reilly, Ormiston, Ortega, O'Shaughnessy, Ossokine, Ottaway, Overmier, Owen, Pace, Page, Page, Pai, Pai, Palamos, Palashov, Palomba, Pal-Singh, Pan, Pan, Pang, Pang, Pankow, Pannarale, Pant, Paoletti, Paoli, Papa, Parida, Parker, Pascucci, Pasqualetti, Passaquieti, Passuello, Patil, Patricelli, Pearlstone, Pedraza, Pedurand, Pekowsky, Pele, Penn, Perez, Perreca, Perri, Pfeiffer, Phelps, Piccinni, Pichot, Piergiovanni, Pierro, Pillant, Pinard, Pinto, Pirello, Pitkin, Poe, Poggiani, Popolizio, Porter, Post, Powell, Prasad, Pratt, Pratten, Predoi, Prestegard, Prijatelj,
  Principe, Privitera, Prix, Prodi, Prokhorov, Puncken, Punturo, Puppo, P\"urrer, Qi, Quetschke, Quintero, Quitzow-James, Raab, Rabeling, Radkins, Raffai, Raja, Rajan, Rajbhandari, Rakhmanov, Ramirez, Ramos-Buades, Rapagnani, Raymond, Razzano, Read, Regimbau, Rei, Reid, Reitze, Ren, Reyes, Ricci, Ricker, Rieger, Riles, Rizzo, Robertson, Robie, Robinet, Rocchi, Rolland, Rollins, Roma, Romano, Romano, Romel, Romie, Rosi\ifmmode~\acute{n}\else \'{n}\fi{}ska, Ross, Rowan, R\"udiger, Ruggi, Rutins, Ryan, Sachdev, Sadecki, Sadeghian, Sakellariadou, Salconi, Saleem, Salemi, Samajdar, Sammut, Sampson, Sanchez, Sanchez, Sanchis-Gual, Sandberg, Sanders, Sassolas, Sathyaprakash, Saulson, Sauter, Savage, Sawadsky, Schale, Scheel, Scheuer, Schmidt, Schmidt, Schnabel, Schofield, Sch\"onbeck, Schreiber, Schuette, Schulte, Schutz, Schwalbe, Scott, Scott, Seidel, Sellers, Sengupta, Sentenac, Sequino, Sergeev, Shaddock, Shaffer, Shah, Shahriar, Shaner, Shao, Shapiro, Shawhan, Sheperd, Shoemaker, Shoemaker, Siellez, Siemens,
  Sieniawska, Sigg, Silva, Singer, Singh, Singhal, Sintes, Slagmolen, Smith, Smith, Smith, Somala, Son, Sonnenberg, Sorazu, Sorrentino, Souradeep, Spencer, Srivastava, Staats, Staley, Steinke, Steinlechner, Steinlechner, Steinmeyer, Stevenson, Stone, Stops, Strain, Stratta, Strigin, Strunk, Sturani, Stuver, Summerscales, Sun, Sunil, Suresh, Sutton, Swinkels, Szczepa\ifmmode~\acute{n}\else \'{n}\fi{}czyk, Tacca, Tait, Talbot, Talukder, Tanner, T\'apai, Taracchini, Tasson, Taylor, Taylor, Tewari, Theeg, Thies, Thomas, Thomas, Thomas, Thorne, Thorne, Thrane, Tiwari, Tiwari, Tokmakov, Toland, Tonelli, Tornasi, Torres-Forn\'e, Torrie, T\"oyr\"a, Travasso, Traylor, Trinastic, Tringali, Trozzo, Tsang, Tse, Tso, Tsukada, Tsuna, Tuyenbayev, Ueno, Ugolini, Unnikrishnan, Urban, Usman, Vahlbruch, Vajente, Valdes, Vallisneri, van Bakel, van Beuzekom, van~den Brand, Van Den~Broeck, Vander-Hyde, van~der Schaaf, van Heijningen, van Veggel, Vardaro, Varma, Vass, Vas\'uth, Vecchio, Vedovato, Veitch, Veitch, Venkateswara,
  Venugopalan, Verkindt, Vetrano, Vicer\'e, Viets, Vinciguerra, Vine, Vinet, Vitale, Vo, Vocca, Vorvick, Vyatchanin, Wade, Wade, Wade, Walet, Walker, Wallace, Walsh, Wang, Wang, Wang, Wang, Wang, Ward, Warner, Was, Watchi, Weaver, Wei, Weinert, Weinstein, Weiss, Wen, Wessel, We\ss{}els, Westerweck, Westphal, Wette, Whelan, Whitcomb, Whiting, Whittle, Wilken, Williams, Williams, Williamson, Willis, Willke, Wimmer, Winkler, Wipf, Wittel, Woan, Woehler, Wofford, Wong, Worden, Wright, Wu, Wysocki, Xiao, Yamamoto, Yancey, Yang, Yap, Yazback, Yu, Yu, Yvert, Zadro\ifmmode~\dot{z}\else \.{z}\fi{}ny, Zanolin, Zelenova, Zendri, Zevin, Zhang, Zhang, Zhang, Zhang, Zhao, Zhou, Zhou, Zhu, Zhu, Zimmerman, Zucker, \& Zweizig}]{Abbott2017PhysRevLett}
Abbott, B.~P., Abbott, R., Abbott, T.~D., {et~al.} 2017, Phys. Rev. Lett., 119, 161101, \dodoi{10.1103/PhysRevLett.119.161101}

\bibitem[{{Beniamini} {et~al.}(2020){Beniamini}, {Granot}, \& {Gill}}]{Beniamini2020MNRAS}
{Beniamini}, P., {Granot}, J., \& {Gill}, R. 2020, \mnras, 493, 3521, \dodoi{10.1093/mnras/staa538}

\bibitem[{{Buckley} {et~al.}(2021){Buckley}, {Bagnulo}, {Britto}, {Mao}, {Kann}, {Cooper}, {Lipunov}, {Hewitt}, {Razzaque}, {Kuin}, {Monageng}, {Covino}, {Jakobsson}, {van der Horst}, {Wiersema}, {B{\"o}ttcher}, {Campana}, {D'Elia}, {Gorbovskoy}, {Gorbunov}, {Groenewald}, {Hartmann}, {Kornilov}, {Mundell}, {Podesta}, {Thomas}, {Tyurina}, {Vlasenko}, {van Soelen}, \& {Xu}}]{Buckley2021MNRAS}
{Buckley}, D.~A.~H., {Bagnulo}, S., {Britto}, R.~J., {et~al.} 2021, \mnras, 506, 4621, \dodoi{10.1093/mnras/stab1791}

\bibitem[{{de Wet} {et~al.}(2023){de Wet}, {Laskar}, {Groot}, {Cavallaro}, {Nicuesa Guelbenzu}, {Chastain}, {Izzo}, {Levan}, {Malesani}, {Monageng}, {van der Horst}, {Zheng}, {Bloemen}, {Filippenko}, {Kann}, {Klose}, {Pieterse}, {Rau}, {Vreeswijk}, {Woudt}, \& {Zhu}}]{deWet2023A&A}
{de Wet}, S., {Laskar}, T., {Groot}, P.~J., {et~al.} 2023, \aap, 671, A116, \dodoi{10.1051/0004-6361/202244917}

\bibitem[{{Dichiara} {et~al.}(2022){Dichiara}, {Troja}, {Lipunov}, {Ricci}, {Oates}, {Butler}, {Liuzzo}, {Ryan}, {O'Connor}, {Cenko}, {Cosentino}, {Lien}, {Gorbovskoy}, {Tyurina}, {Balanutsa}, {Vlasenko}, {Gorbunov}, {Podesta}, {Podesta}, {Rebolo}, {Serra}, \& {Buckley}}]{Dichiara2022MNRAS}
{Dichiara}, S., {Troja}, E., {Lipunov}, V., {et~al.} 2022, \mnras, 512, 2337, \dodoi{10.1093/mnras/stac454}

\bibitem[{{Eichler} {et~al.}(1989){Eichler}, {Livio}, {Piran}, \& {Schramm}}]{Eichler1989Nature}
{Eichler}, D., {Livio}, M., {Piran}, T., \& {Schramm}, D.~N. 1989, \nat, 340, 126, \dodoi{10.1038/340126a0}

\bibitem[{{Gao} {et~al.}(2013){Gao}, {Lei}, {Wu}, \& {Zhang}}]{Gao2013MNRAS}
{Gao}, H., {Lei}, W.-H., {Wu}, X.-F., \& {Zhang}, B. 2013, \mnras, 435, 2520, \dodoi{10.1093/mnras/stt1461}

\bibitem[{{Ghisellini} {et~al.}(1998){Ghisellini}, {Haardt}, \& {Svensson}}]{Ghisellini1998MNRAS}
{Ghisellini}, G., {Haardt}, F., \& {Svensson}, R. 1998, \mnras, 297, 348, \dodoi{10.1046/j.1365-8711.1998.01442.x}

\bibitem[{{Ghisellini} \& {Lazzati}(1999)}]{Ghisellini1999MNRAS}
{Ghisellini}, G., \& {Lazzati}, D. 1999, \mnras, 309, L7, \dodoi{10.1046/j.1365-8711.1999.03025.x}

\bibitem[{{Gill} {et~al.}(2020){Gill}, {Granot}, \& {Kumar}}]{Gill2020MNRAS}
{Gill}, R., {Granot}, J., \& {Kumar}, P. 2020, \mnras, 491, 3343, \dodoi{10.1093/mnras/stz2976}

\bibitem[{{Granot} {et~al.}(2002){Granot}, {Panaitescu}, {Kumar}, \& {Woosley}}]{Granot2002ApJL}
{Granot}, J., {Panaitescu}, A., {Kumar}, P., \& {Woosley}, S.~E. 2002, \apjl, 570, L61, \dodoi{10.1086/340991}

\bibitem[{{Granot} \& {Taylor}(2005)}]{Granot2005ApJ}
{Granot}, J., \& {Taylor}, G.~B. 2005, \apj, 625, 263, \dodoi{10.1086/429536}

\bibitem[{{Huang} {et~al.}(2019){Huang}, {Lin}, {Liu}, {Ren}, {Wang}, {Liu}, \& {Liang}}]{Huang2019MNRAS}
{Huang}, B.-Q., {Lin}, D.-B., {Liu}, T., {et~al.} 2019, \mnras, 487, 3214, \dodoi{10.1093/mnras/stz1426}

\bibitem[{{Huang} {et~al.}(1999{\natexlab{a}}){Huang}, {Dai}, \& {Lu}}]{Huang1999ChPhL}
{Huang}, Y.-f., {Dai}, Z.-g., \& {Lu}, T. 1999{\natexlab{a}}, Chinese Physics Letters, 16, 775, \dodoi{10.1088/0256-307X/16/10/027}

\bibitem[{{Huang} {et~al.}(1999{\natexlab{b}}){Huang}, {Dai}, \& {Lu}}]{Huang1999MNRAS}
{Huang}, Y.~F., {Dai}, Z.~G., \& {Lu}, T. 1999{\natexlab{b}}, \mnras, 309, 513, \dodoi{10.1046/j.1365-8711.1999.02887.x}

\bibitem[{{Huang} {et~al.}(2000){Huang}, {Gou}, {Dai}, \& {Lu}}]{Huang2000ApJ}
{Huang}, Y.~F., {Gou}, L.~J., {Dai}, Z.~G., \& {Lu}, T. 2000, \apj, 543, 90, \dodoi{10.1086/317076}

\bibitem[{{Huang} {et~al.}(2007){Huang}, {Lu}, {Wong}, \& {Cheng}}]{Huang2007ChJAA}
{Huang}, Y.-F., {Lu}, Y., {Wong}, A. Y.~L., \& {Cheng}, K.~S. 2007, \cjaa, 7, 397, \dodoi{10.1088/1009-9271/7/3/09}

\bibitem[{{Huang} {et~al.}(2004){Huang}, {Wu}, {Dai}, {Ma}, \& {Lu}}]{Huang2004ApJ}
{Huang}, Y.~F., {Wu}, X.~F., {Dai}, Z.~G., {Ma}, H.~T., \& {Lu}, T. 2004, \apj, 605, 300, \dodoi{10.1086/382202}

\bibitem[{{Ioka} {et~al.}(2005){Ioka}, {Kobayashi}, \& {Zhang}}]{Ioka2005ApJ}
{Ioka}, K., {Kobayashi}, S., \& {Zhang}, B. 2005, \apj, 631, 429, \dodoi{10.1086/432567}

\bibitem[{Laing(1980)}]{Laing1980MNRAS}
Laing, R.~A. 1980, Monthly Notices of the Royal Astronomical Society, 193, 439, \dodoi{10.1093/mnras/193.3.439}

\bibitem[{{Lamb} {et~al.}(2022){Lamb}, {Nativi}, {Rosswog}, {Kann}, {Levan}, {Lundman}, \& {Tanvir}}]{Lamb2022Univ}
{Lamb}, G.~P., {Nativi}, L., {Rosswog}, S., {et~al.} 2022, Universe, 8, 612, \dodoi{10.3390/universe8120612}

\bibitem[{{Lan} {et~al.}(2016){Lan}, {Wu}, \& {Dai}}]{Lan2016ApJL}
{Lan}, M.-X., {Wu}, X.-F., \& {Dai}, Z.-G. 2016, \apj, 816, 73, \dodoi{10.3847/0004-637X/816/2/73}

\bibitem[{{Lan} {et~al.}(2023){Lan}, {Wu}, \& {Dai}}]{Lan2023ApJL}
---. 2023, \apj, 952, 31, \dodoi{10.3847/1538-4357/acd6ef}

\bibitem[{{Laskar} {et~al.}(2019){Laskar}, {Alexander}, {Gill}, {Granot}, {Berger}, {Mundell}, {Barniol Duran}, {Bolmer}, {Duffell}, {van Eerten}, {Fong}, {Kobayashi}, {Margutti}, \& {Schady}}]{Laskar2019ApJ}
{Laskar}, T., {Alexander}, K.~D., {Gill}, R., {et~al.} 2019, \apjl, 878, L26, \dodoi{10.3847/2041-8213/ab2247}

\bibitem[{{Li} {et~al.}(2023){Li}, {Gao}, {Ai}, \& {Lei}}]{Li2023MNRAS}
{Li}, J.-D., {Gao}, H., {Ai}, S., \& {Lei}, W.-H. 2023, \mnras, 525, 6285, \dodoi{10.1093/mnras/stad2606}

\bibitem[{{Lipunov} {et~al.}(2019){Lipunov}, {Vladimirov}, {Gorbovskoi}, {Kuznetsov}, {Zimnukhov}, {Balanutsa}, {Kornilov}, {Tyurina}, {Gress}, {Vlasenko}, {Gabovich}, {Yurkov}, {Kuvshinov}, \& {Senik}}]{Lipunov2019ARep}
{Lipunov}, V.~M., {Vladimirov}, V.~V., {Gorbovskoi}, E.~S., {et~al.} 2019, Astronomy Reports, 63, 293, \dodoi{10.1134/S1063772919040073}

\bibitem[{{Longair}(1994)}]{Longair1994}
{Longair}, M.~S. 1994, {High energy astrophysics. Vol.2: Stars, the galaxy and the interstellar medium}, Vol.~2

\bibitem[{{MacFadyen} \& {Woosley}(1999)}]{MacFadyen1999APJ}
{MacFadyen}, A.~I., \& {Woosley}, S.~E. 1999, \apj, 524, 262, \dodoi{10.1086/307790}

\bibitem[{{Mao} {et~al.}(2018){Mao}, {Covino}, \& {Wang}}]{Mao2018ApJ}
{Mao}, J., {Covino}, S., \& {Wang}, J. 2018, \apj, 860, 153, \dodoi{10.3847/1538-4357/aac5d9}

\bibitem[{{Medvedev} \& {Loeb}(1999)}]{Medvedev1999ApJ}
{Medvedev}, M.~V., \& {Loeb}, A. 1999, \apj, 526, 697, \dodoi{10.1086/308038}

\bibitem[{{M{\'e}sz{\'a}ros} \& {Rees}(1997)}]{MeszarosRees1997ApJ}
{M{\'e}sz{\'a}ros}, P., \& {Rees}, M.~J. 1997, \apjl, 482, L29, \dodoi{10.1086/310692}

\bibitem[{{M{\'e}sz{\'a}ros} {et~al.}(1998){M{\'e}sz{\'a}ros}, {Rees}, \& {Wijers}}]{Meszaros1998APJ}
{M{\'e}sz{\'a}ros}, P., {Rees}, M.~J., \& {Wijers}, R.~A.~M.~J. 1998, \apj, 499, 301, \dodoi{10.1086/305635}

\bibitem[{{Mundell} {et~al.}(2013){Mundell}, {Kopa{\v{c}}}, {Arnold}, {Steele}, {Gomboc}, {Kobayashi}, {Harrison}, {Smith}, {Guidorzi}, {Virgili}, {Melandri}, \& {Japelj}}]{Mundell2013Nature}
{Mundell}, C.~G., {Kopa{\v{c}}}, D., {Arnold}, D.~M., {et~al.} 2013, \nat, 504, 119, \dodoi{10.1038/nature12814}

\bibitem[{{Nakamura}(2000)}]{Nakamura2000ApJL}
{Nakamura}, T. 2000, \apjl, 534, L159, \dodoi{10.1086/312663}

\bibitem[{{Nakar} {et~al.}(2009){Nakar}, {Ando}, \& {Sari}}]{Nakar2009ApJ}
{Nakar}, E., {Ando}, S., \& {Sari}, R. 2009, \apj, 703, 675, \dodoi{10.1088/0004-637X/703/1/675}

\bibitem[{{Narayan} \& {Kumar}(2009)}]{Narayan2009MNRAS}
{Narayan}, R., \& {Kumar}, P. 2009, \mnras, 394, L117, \dodoi{10.1111/j.1745-3933.2009.00624.x}

\bibitem[{{Narayan} {et~al.}(1992){Narayan}, {Paczynski}, \& {Piran}}]{Narayan1992APJL}
{Narayan}, R., {Paczynski}, B., \& {Piran}, T. 1992, \apjl, 395, L83, \dodoi{10.1086/186493}

\bibitem[{{ORSI}(2011)}]{ORSI2011ICRC}
{ORSI}, S. 2011, in International Cosmic Ray Conference, Vol.~6, International Cosmic Ray Conference, 403, \dodoi{10.7529/ICRC2011/V06/1128}

\bibitem[{{Orsi} \& {Polar Collaboration}(2011)}]{Orsi2011ASTRA}
{Orsi}, S., \& {Polar Collaboration}. 2011, Astrophysics and Space Sciences Transactions, 7, 43, \dodoi{10.5194/astra-7-43-2011}

\bibitem[{{Paczynski}(1986)}]{Paczynski1986APJL}
{Paczynski}, B. 1986, \apjl, 308, L43, \dodoi{10.1086/184740}

\bibitem[{{Paczy{\'n}ski}(1991)}]{Paczynski1991bGRB}
{Paczy{\'n}ski}, B. 1991, in American Institute of Physics Conference Series, Vol. 265, Gamma-ray Bursts, 144--148, \dodoi{10.1063/1.42815}

\bibitem[{{Paczynski}(1991)}]{Paczynski1991aACTAA}
---. 1991, \actaa, 41, 257

\bibitem[{{Paczy{\'n}ski}(1998)}]{Paczynski1998}
{Paczy{\'n}ski}, B. 1998, in American Institute of Physics Conference Series, Vol. 428, Gamma-Ray Bursts, 4th Hunstville Symposium, ed. C.~A. {Meegan}, R.~D. {Preece}, \& T.~M. {Koshut}, 783--787, \dodoi{10.1063/1.55404}

\bibitem[{{Peng} {et~al.}(2005){Peng}, {K{\"o}nigl}, \& {Granot}}]{Peng2005ApJ}
{Peng}, F., {K{\"o}nigl}, A., \& {Granot}, J. 2005, \apj, 626, 966, \dodoi{10.1086/430045}

\bibitem[{{Rees} \& {Meszaros}(1994)}]{Rees1994ApJ}
{Rees}, M.~J., \& {Meszaros}, P. 1994, \apjl, 430, L93, \dodoi{10.1086/187446}

\bibitem[{{Rossi} {et~al.}(2004){Rossi}, {Lazzati}, {Salmonson}, \& {Ghisellini}}]{Rossi2004MNRAS}
{Rossi}, E.~M., {Lazzati}, D., {Salmonson}, J.~D., \& {Ghisellini}, G. 2004, \mnras, 354, 86, \dodoi{10.1111/j.1365-2966.2004.08165.x}

\bibitem[{{Sari}(1999)}]{Sari1999ApJ}
{Sari}, R. 1999, \apjl, 524, L43, \dodoi{10.1086/312294}

\bibitem[{{Sari} \& {Esin}(2001)}]{Sari2001ApJ}
{Sari}, R., \& {Esin}, A.~A. 2001, \apj, 548, 787, \dodoi{10.1086/319003}

\bibitem[{{Sari} {et~al.}(1998){Sari}, {Piran}, \& {Narayan}}]{Sari1998ApJ}
{Sari}, R., {Piran}, T., \& {Narayan}, R. 1998, \apjl, 497, L17, \dodoi{10.1086/311269}

\bibitem[{{Shrestha} {et~al.}(2020){Shrestha}, {Steele}, {Piascik}, {Jermak}, {Smith}, \& {Copperwheat}}]{Shrestha2020MNRAS}
{Shrestha}, M., {Steele}, I.~A., {Piascik}, A.~S., {et~al.} 2020, \mnras, 494, 4676, \dodoi{10.1093/mnras/staa1049}

\bibitem[{{Steele} {et~al.}(2009){Steele}, {Mundell}, {Smith}, {Kobayashi}, \& {Guidorzi}}]{Steele2009Nature}
{Steele}, I.~A., {Mundell}, C.~G., {Smith}, R.~J., {Kobayashi}, S., \& {Guidorzi}, C. 2009, \nat, 462, 767, \dodoi{10.1038/nature08590}

\bibitem[{{Urata} {et~al.}(2019){Urata}, {Toma}, {Huang}, {Asada}, {Nagai}, {Takahashi}, {Petitpas}, {Tashiro}, \& {Yamaoka}}]{Urata2019ApJ}
{Urata}, Y., {Toma}, K., {Huang}, K., {et~al.} 2019, \apjl, 884, L58, \dodoi{10.3847/2041-8213/ab48f3}

\bibitem[{{Urata} {et~al.}(2023){Urata}, {Toma}, {Covino}, {Wiersema}, {Huang}, {Shimoda}, {Kuwata}, {Nagao}, {Asada}, {Nagai}, {Takahashi}, {Chung}, {Petitpas}, {Yamaoka}, {Izzo}, {Fynbo}, {de Ugarte Postigo}, {Arabsalmani}, \& {Tashiro}}]{Urata2023NatAs}
{Urata}, Y., {Toma}, K., {Covino}, S., {et~al.} 2023, Nature Astronomy, 7, 80, \dodoi{10.1038/s41550-022-01832-7}

\bibitem[{{Warren} {et~al.}(2018){Warren}, {Barkov}, {Ito}, {Nagataki}, \& {Laskar}}]{Warren2018MNRAS}
{Warren}, D.~C., {Barkov}, M.~V., {Ito}, H., {Nagataki}, S., \& {Laskar}, T. 2018, \mnras, 480, 4060, \dodoi{10.1093/mnras/sty2138}

\bibitem[{{Woosley}(1993)}]{Woosley1993APJ}
{Woosley}, S.~E. 1993, \apj, 405, 273, \dodoi{10.1086/172359}

\bibitem[{{Woosley} \& {Bloom}(2006)}]{Woosley2006ARAA}
{Woosley}, S.~E., \& {Bloom}, J.~S. 2006, \araa, 44, 507, \dodoi{10.1146/annurev.astro.43.072103.150558}

\bibitem[{{Wu} {et~al.}(2003){Wu}, {Dai}, {Huang}, \& {Lu}}]{Wu2003MNRAS}
{Wu}, X.~F., {Dai}, Z.~G., {Huang}, Y.~F., \& {Lu}, T. 2003, \mnras, 342, 1131, \dodoi{10.1046/j.1365-8711.2003.06602.x}

\bibitem[{{Wu} {et~al.}(2005){Wu}, {Dai}, {Huang}, \& {Lu}}]{Wu2005MNRAS}
---. 2005, \mnras, 357, 1197, \dodoi{10.1111/j.1365-2966.2005.08685.x}

\bibitem[{{Yamazaki} {et~al.}(2004){Yamazaki}, {Ioka}, \& {Nakamura}}]{Yamazaki2004APJL}
{Yamazaki}, R., {Ioka}, K., \& {Nakamura}, T. 2004, \apjl, 607, L103, \dodoi{10.1086/421872}

\bibitem[{{Zhang}(2018)}]{2018pgrb.book}
{Zhang}, B. 2018, {The Physics of Gamma-Ray Bursts}, \dodoi{10.1017/9781139226530}

\bibitem[{{Zhang} \& {Yan}(2011)}]{ZhangB2011ApJ}
{Zhang}, B., \& {Yan}, H. 2011, \apj, 726, 90, \dodoi{10.1088/0004-637X/726/2/90}

\end{thebibliography}
\bibliographystyle{aasjournal}

%% This command is needed to show the entire author+affiliation list when
%% the collaboration and author truncation commands are used.  It has to
%% go at the end of the manuscript.
%\allauthors

%% Include this line if you are using the \added, \replaced, \deleted
%% commands to see a summary list of all changes at the end of the article.
%\listofchanges

\appendix
The polarization of afterglow is sensitive to the shape of the light curve and the spectrum. \citet{Li2023MNRAS} have developed a method for calculating the afterglow of GRB, which obtains physical parameters at each moment through numerical dynamic evolution and uses these parameters to analytically calculate the radiation flux of afterglow. On the basis of \citet{Li2023MNRAS}, we added SSC radiation and further adjusted the algorithm to make it more suitable for polarization calculation (see Section \ref{math} for details). We refer to the code used in this work as AFGoLipy.

\section{Synchrotron Radiation}
In the case of the $i-$th element shocking the interstellar medium, the characteristic synchrotron frequency of electrons with Lorentz factor $\gamma_e>>1$ in magnetic field $B$ in the co-moving frame is \citep{Sari1998ApJ}:
\begin{equation}   \nu\left(\gamma_e\right)=\gamma\gamma_e^2\frac{q_eB}{2\pi m_ec},
   \label{eq:nu_char}
\end{equation}
where $m_e$, $q_e$ are the mass and charge of a electron, respectively. $c$ is the speed of light. $\gamma$ is the Lorentz factor of the element’s bulk motion, whose evolution can be expressed as \citep{Huang1999ChPhL,Huang1999MNRAS}:
\begin{equation}
   \frac{d\gamma}{dm}=-\frac{\gamma^2-1}{M_{\text{ej}}+\varepsilon m+2(1-\varepsilon)\gamma m},
   \label{eq:dgammma_dm}
\end{equation}
where $M_{\text{ej}}=E_0/\left(\gamma_0c^2\right)$ is the ejecta mass, and $E_0$ is the initial kinetic energy of an element. The $\varepsilon$ is radiative efficiency, which is defined as the proportion of the internal energy generated by the shock in jet's comoving frame that would be radiated. $m$ is the mass of the interstellar medium swept by the element. If an element occupies the azimuth range from $\varphi_{\text{l}}$ to $\varphi_{\text{u}}$ on the cross-section of the jet, the evolution of $m$ can be written as:
\begin{equation}
    \frac{dm}{dR}=\left(\varphi_{\text{u}}-\varphi_{\text{l}}\right)R^2\left(1-\cos{\theta_{\text{j}}}\right)nm_{\text{p}},
\end{equation}
where $m_p$ is the mass of a proton. $n$ is the particle number density of interstellar medium, which can be described as $AR^{-k}$, where $k$ is the wind profile variable. And $R$ is the jet's radius. For the uniform interstellar medium, $k=0$. And for the stellar wind environment, $k=2$. $A$ is a constant. The evolution of the jet's radius $R$ over time $t$ in the frame of an on-axis observer can reads as: 
\begin{equation}
    \frac{dR}{dt}=\beta c\gamma(\gamma +\sqrt{\gamma^2-1}).
    \label{eq:dR_dT}
\end{equation}
The $\beta$ represent the dimensionless velocity of the element's bulk motion.

The minimum Lorentz factor for the random motion of electrons can be derived as \citep{Huang2000ApJ}
\begin{equation}
\gamma_{e,\text{min}}=\epsilon_e\left(\gamma-1\right)\frac{m_p\left(p-2\right)}{m_e\left(p-1\right)}+1,
\end{equation} 
where $\epsilon_e$ is the fraction of the total shock generated internal energy goes into the electrons. $p$ represents the power-law distribution index of electrons in interstellar media ($\text{d}N_e/\text{d}\gamma_e \propto \gamma_e^{-p}$). And the characteristic Lorentz factor $\gamma_c$ is defined as \citep{Sari1998ApJ}:
\begin{equation}
    \gamma_c=\frac{6\pi m_ec}{\sigma_TB^2t}=\frac{3m_e}{16\epsilon_B\sigma_Tm_pc}\frac{1}{t\gamma^3n},
\end{equation}
beyond which the electrons might have significantly cooled. Where $\sigma_T$ is the the cross section for Thompson scattering. If the Synchrotron Self-Compton (SSC) scattering is significant, the cooling of electrons is enhanced. Therefore the $\gamma_c$ should to be divided by $(1+Y)$, where $Y$ is the Compton $y$-parameter. The $\gamma_{e,\text{min}}$, $\gamma_c$ defines the characteristic frequencies $\nu_m$ and $\nu_c$, which divide afterglow radiation into two situations: $\nu_m>\nu_c$ for fast cooling and $\nu_m<\nu_c$ for slow cooling.

In an $n=AR^{-k}$ environment, for the slow cooling regime ($\nu_c>\nu_m$), the self absorption frequency $\nu_a$ is
\begin{equation}
\nu_a=\begin{cases}
\left[\frac{c_1q_enR}{\left(3-k\right)B\gamma_c^5}\right]^{3/5}\nu_m &\nu_a<\nu_m,\\
\left[\frac{c_2q_enR}{\left(3-k\right)B\gamma_c^5}\right]^{2/\left(p+4\right)}\nu_m &\nu_m<\nu_a<\nu_c,\\
\left[\frac{c_2q_enR}{\left(3-k\right)B\gamma_c^5}\right]^{2/\left(p+5\right)}\left(\frac{\nu_c}{\nu_m}\right)^{1/\left(p+5\right)}\nu_m &\nu_c<\nu_a.
\end{cases}
\end{equation}
$c_1$ and $c_2$ are coefficients dependent on $p$ \citep{Wu2003MNRAS}. Assuming a point source of the element is on the LOS, the observed flux density $F_{\nu}$ is divided into the following three situations\\
(1)$\nu_a<\nu_m<\nu_c$: 
\begin{equation}
\label{lc1}
F_{\nu}^{\text{syn}}=\frac{1}{\Omega}F_{\nu,\text{max}}\begin{cases}
\left(\frac{\nu}{\nu_a}\right)^2\left(\frac{\nu_a}{\nu_m}\right)^{1/3} &\nu<\nu_a,\\
\left(\frac{\nu}{\nu_m}\right)^{1/3} &\nu_a<\nu<\nu_m,\\
\left(\frac{\nu}{\nu_m}\right)^{-\left(p-1\right)/2}F_{\nu,\text{max}} &\nu_m<\nu<\nu_c,\\
\left(\frac{\nu}{\nu_m}\right)^{-\left(p-1\right)/2}\left(\frac{\nu}{\nu_c}\right)^{-p/2} &\nu_c<\nu<\nu_M.
\end{cases}
\end{equation}
(2)$\nu_m<\nu_a<\nu_c$: 
\begin{equation}
\label{lc2}
F_{\nu}^{\text{syn}}=\frac{1}{\Omega}F_{\nu,\text{max}}\begin{cases}
\left(\frac{\nu}{\nu_m}\right)^2\left(\frac{\nu_m}{\nu_a}\right)^{\left(p+4\right)/2}F_{\nu,\text{max}} &\nu<\nu_m,\\
\left(\frac{\nu}{\nu_a}\right)^{5/2}\left(\frac{\nu_a}{\nu_m}\right)^{-\left(p-1\right)/2}F_{\nu,\text{max}} &\nu_m<\nu<\nu_a,\\
\left(\frac{\nu}{\nu_m}\right)^{-\left(p-1\right)/2}F_{\nu,\text{max}} &\nu_a<\nu<\nu_c,\\
\left(\frac{\nu}{\nu_m}\right)^{-\left(p-1\right)/2}\left(\frac{\nu}{\nu_c}\right)^{-p/2}F_{\nu,\text{max}} &\nu_c<\nu<\nu_M.
\end{cases}
\end{equation}
(3)$\nu_m<\nu_c<\nu_a$: 
\begin{equation}
\label{lc3}
F_{\nu}^{\text{syn}}=\frac{1}{\Omega}F_{\nu,\text{max}}\begin{cases}
\left(\frac{\nu}{\nu_m}\right)^2\left(\frac{\nu_m}{\nu_a}\right)^{\left(p+4\right)/2}\left(\frac{\nu_a}{\nu_c}\right)^{-1/2} &\nu<\nu_m,\\
\left(\frac{\nu}{\nu_a}\right)^{5/2}\left(\frac{\nu_a}{\nu_c}\right)^{-p/2}\left(\frac{\nu_c}{\nu_m}\right)^{-\left(p-1\right)/2} &\nu_m<\nu<\nu_a,\\
\left(\frac{\nu}{\nu_c}\right)^{-p/2}\left(\frac{\nu_c}{\nu_m}\right)^{-\left(p-1\right)/2} &\nu_a<\nu<\nu_M.
\end{cases}
\end{equation}
where $F_{\nu,\text{max}}$ stands for the peak flux density of the spectrum. $\nu_M$ is the maximum frequency of synchrotron radiation, which is defined by the maximum Lorentz factor of electrons $\gamma_M$. The parameter $\Omega$ is the solid angle of the beaming cone of radiation due to the relativistic beaming effect of the point source, which can be expressed as $\Omega=2\pi\left[1-\cos{\left(1/\gamma\right)}\right]$, where $1/\gamma$ is the half-opening angle of the beaming cone. To obtain the flux from the point source, introducing the factor $1/\Omega$ is necessary. Incidentally, although the polarization below the synchrotron self-absorption frequency is suppressed, we still present the radiation flux below this frequency for the sake of completeness. In the fast cooling regime($\nu_c<\nu_m$), the self absorption frequency $\nu_a$ is:
\begin{equation}
\nu_a=\begin{cases}
\left[\frac{c_1q_enR}{\left(3-k\right)B\gamma_c^5}\right]^{3/5}\nu_c &\nu_a<\nu_c,\\
\left[\frac{c_2q_enR}{\left(3-k\right)B\gamma_c^5}\right]^{1/3}\nu_c &\nu_c<\nu_a<\nu_m,\\
\left[\frac{c_2q_enR}{\left(3-k\right)B\gamma_c^5}\right]^{2/\left(p+5\right)}\left(\frac{\nu_m}{\nu_c}\right)^{\left(p-1\right)/\left(p+5\right)}\nu_c &\nu_m<\nu_a.
\end{cases}
\end{equation}
And the flux in fast cooling regime is\\
(1)$\nu_a<\nu_c<\nu_m$: 
\begin{equation}
\label{fc1}
F_{\nu}^{\text{syn}}=\frac{1}{\Omega}F_{\nu,\text{max}}\begin{cases}
\left(\frac{\nu}{\nu_a}\right)^2\left(\frac{\nu_a}{\nu_c}\right)^{1/3} &\nu<\nu_a,\\
\left(\frac{\nu}{\nu_c}\right)^{1/3} &\nu_a<\nu<\nu_c,\\
\left(\frac{\nu}{\nu_c}\right)^{-1/2} &\nu_c<\nu<\nu_m,\\
\left(\frac{\nu_m}{\nu_c}\right)^{-1/2}\left(\frac{\nu}{\nu_m}\right)^{-p/2} &\nu_m<\nu<\nu_M.
\end{cases}
\end{equation}
(2)$\nu_c<\nu_a<\nu_m$: 
\begin{equation}
\label{fc2}
F_{\nu}^{\text{syn}}=\frac{1}{\Omega}F_{\nu,\text{max}}\begin{cases}
\left(\frac{\nu}{\nu_c}\right)^2\left(\frac{\nu_c}{\nu_a}\right)^3 &\nu<\nu_c,\\
\left(\frac{\nu}{\nu_a}\right)^{5/2}\left(\frac{\nu_a}{\nu_c}\right)^{-1/2} &\nu_c<\nu<\nu_a,\\
\left(\frac{\nu}{\nu_c}\right)^{-1/2} &\nu_a<\nu<\nu_m,\\
\left(\frac{\nu_m}{\nu_c}\right)^{-1/2}\left(\frac{\nu}{\nu_m}\right)^{-p/2} &\nu_m<\nu<\nu_M.
\end{cases}
\end{equation}
(3)$\nu_c<\nu_m<\nu_a$: 
\begin{equation}
\label{fc3}
F_{\nu}^{\text{syn}}=\frac{1}{\Omega}F_{\nu,\text{max}}\begin{cases}
\left(\frac{\nu}{\nu_c}\right)^2\left(\frac{\nu_c}{\nu_a}\right)^3\left(\frac{\nu_a}{\nu_m}\right)^{-\left(p-1\right)/2} &\nu<\nu_c,\\
\left(\frac{\nu}{\nu_a}\right)^{5/2}\left(\frac{\nu_a}{\nu_m}\right)^{-p/2}\left(\frac{\nu_m}{\nu_c}\right)^{-1/2} &\nu_c<\nu<\nu_a,\\
\left(\frac{\nu}{\nu_m}\right)^{-p/2}\left(\frac{\nu_m}{\nu_c}\right)^{-1/2} &\nu_a<\nu<\nu_M.
\end{cases}
\end{equation}

It is worth noting that in cases of strong synchrotron self-absorption regime (max$(\nu_m, \nu_c)<\nu_a$), the dynamical evolution may be influenced by synchrotron self-absorption. For example, the energy distribution of electrons may be closer to a Maxwellian distribution rather than a power-law distribution \citep{Ghisellini1998MNRAS}. Therefore, readers should apply the above equations with caution when dealing with strong synchrotron self-absorption regime.

\section{Synchrotron Self-Compton scattering}
The characteristic frequencies of SSC emission are defined by the different combination of the characteristic frequencies $\nu_{a}$, $\nu_{m}$, $\nu_{c}$ and characteristic Lorentz factor $\gamma_{a}$, $\gamma_{m}$, $\gamma_{c}$ of synchrotron radiation \citep{Sari2001ApJ}. It is convenient for us to define the characteristic frequencies as
\begin{equation}
\nu_{ij}^{\text{IC}}=4\gamma_i^2\nu_jx_0, 
\end{equation}
with the subcripts $i, j=a, c, m$. The characteristic frequencies represent the characteristic upscattered frequency for mono-energetic photons with frequency $\nu_j$ scattered by mono-energetic electrons with Lorentz factor $\gamma_i$. The factor $x_0\sim0.5$ is necessary to ensure energy conservation.

In this section, we assume that only the first-order SSC component is important. When the Klein-Nishina
correction can be neglected, \citet{Gao2013MNRAS} approximate the analytical SSC spectra. Similar to the synchrotron radiation, the SSC spectra is determined by the relationship among $\nu_{a}$, $\nu_{m}$ and $\nu_{c}$:\\
(1)$\nu_{a}<\nu_{m}<\nu_{c}$:
\begin{equation}
F_{\nu}^{\text{IC}}=\frac{1}{\Omega}\tau_{\text{es}}F_{\nu,\text{max}}x_0\\
\times\begin{cases}
\frac{5}{2}\frac{p-1}{p+1}\left(\frac{\nu_{a}}{\nu_{m}}\right)^{\frac{1}{3}}\left(\frac{\nu}{\nu_{ma}^{\text{IC}}}\right) &\nu<\nu_{ma}^{\text{IC}},\\
\frac{3}{2}\frac{p-1}{p-1/3}\left(\frac{\nu}{\nu_{ma}^{\text{IC}}}\right)^{\frac{1}{3}} &\nu_{ma}^{\text{IC}}<\nu<\nu_{mm}^{\text{IC}},\\
\frac{p-1}{p+1}\left(\frac{\nu}{\nu_{mm}^{\text{IC}}}\right)^{\frac{1-p}{2}}\left[\frac{4(p+1/3)}{(p+1)(p-1/3)}+\ln\left(\frac{\nu}{\nu_{mm}^{\text{IC}}}\right)\right] &\nu_{mm}^{\text{IC}}<\nu<\nu_{mc}^{\text{IC}},\\
\frac{p-1}{p+1}\left(\frac{\nu}{\nu_{mm}^{\text{IC}}}\right)^{\frac{1-p}{2}}\left[\frac{2(2p+3)}{p+2}-\frac{2}{(p+1)(p+2)}+\ln\left(\frac{\nu_{cc}^{\text{IC}}}{\nu}\right)\right] &\nu_{mc}^{\text{IC}}<\nu<\nu_{cc}^{\text{IC}},\\
\frac{p-1}{p+1}\left(\frac{\nu}{\nu_{mm}^{\text{IC}}}\right)^{-\frac{p}{2}}\left(\frac{\nu_{c}}{\nu_{m}}\right)\left[\frac{2(2p+3)}{p+2}-\frac{2}{(p+2)^{2}}+\frac{p+1}{p+2}\ln\left(\frac{\nu}{\nu_{cc}^{\text{IC}}}\right)\right] &\nu_{cc}^{\text{IC}}<\nu.
\end{cases}
\end{equation}
(2)$\nu_{m}<\nu_{a}<\nu_{c}$:
\begin{equation}
F_{\nu}^{\text{IC}}=\frac{1}{\Omega}\tau_{\text{es}}F_{\nu,\text{max}}x_0\\
\times\begin{cases}
\frac{2(p+4)(p-1)}{3(p+1)^2}\left(\frac{\nu_\text{m}}{\nu_\text{a}}\right)^{\frac{p+1}{2}}\left(\frac{\nu}{\nu_{mm}^{\text{IC}}}\right) &\nu<\nu_{ma}^{\text{IC}},\\
\frac{p-1}{p+1}\left(\frac{\nu}{\nu_{mm}^{\text{IC}}}\right)^{\frac{1-p}{2}}\left[\frac{2(2p+5)}{(p+1)(p+4)}+\ln\left(\frac{\nu}{\nu_{ma}^{\text{IC}}}\right)\right] &\nu_{ma}^{\text{IC}}<\nu<\nu_{mc}^{\text{IC}},\\
\frac{p-1}{p+1}\left(\frac{\nu}{\nu_{mm}^{\text{IC}}}\right)^{\frac{1-p}{2}}\left[2+\frac{2}{p+4}+\ln\left(\frac{\nu}{\nu_{a}}\right)\right] &\nu_{mc}^{\text{IC}}<\nu<\nu_{ca}^{\text{IC}},\\
\frac{p-1}{p+1}\left(\frac{\nu}{\nu_{mm}^{\text{IC}}}\right)^{\frac{1-p}{2}}\left[\frac{2(2p+1)}{p+1}+\ln\left(\frac{\nu_{cc}^{\text{IC}}}{\nu}\right)\right] &\nu_{ca}^{\text{IC}}<\nu<\nu_{cc}^{\text{IC}},\\
\frac{p-1}{p+2}\left(\frac{\nu_{c}}{\nu_{m}}\right)\left(\frac{\nu}{\nu_{mm}^{\text{IC}}}\right)^{-\frac{p}{2}}\left[\frac{2(2p+5)}{p+2}+\ln\left(\frac{\nu}{\nu_{cc}^{\text{IC}}}\right)\right],&\nu>\nu_{cc}^{\text{IC}}.
\end{cases}
\end{equation}
(3)$\nu_{a}<\nu_{c}<\nu_{m}$:
\begin{equation}
F_{\nu}^{\text{IC}}=\frac{1}{\Omega}\tau_{\text{es}}F_{\nu,\text{max}}x_0\\
\times\begin{cases}
\frac{5}{6}\left(\frac{\nu_{a}}{\nu_{c}}\right)^{\frac{1}{2}}\left(\frac{\nu}{\nu_{ca}}\right) &\nu<\nu_{ca}^{\text{IC}},\\
\frac{9}{10}\left(\frac{\nu}{\nu_{cc}^{{\text{IC}}}}\right)^{{\frac{1}{3}}} &\nu_{ca}^{{\text{IC}}}<\nu<\nu_{cc}^{{\text{IC}}},\\
\frac{1}{3}\left(\frac{\nu}{\nu_{cc}^{{\text{IC}}}}\right)^{{-\frac{1}{2}}}\left[\frac{28}{15}+\ln\left(\frac{\nu}{\nu_{cc}^{{\text{IC}}}}\right)\right] &\nu_{cc}^{{\text{IC}}}<\nu<\nu_{cm}^{{\text{IC}}},\\
\frac{1}{3}\left(\frac{\nu}{\nu_{cc}^{{\text{IC}}}}\right)^{{-\frac{1}{2}}}\left[\frac{2(p+5)}{(p+2)(p-1)}-\frac{2(p-1)}{3(p+2)}+\ln\left(\frac{\nu_{mm}^{{\text{IC}}}}{\nu}\right)\right] &\nu_{cm}^{{\text{IC}}}<\nu<\nu_{mm}^{{\text{IC}}},\\
\frac{1}{p+2}\left(\frac{\nu_{c}}{\nu_{m}}\right)\left(\frac{\nu}{\nu_{mm}^{{\text{IC}}}}\right)^{{-\frac{p}{2}}}\left[\frac{2}{3}\frac{p+5}{p-1}-\frac{2}{3}\frac{p-1}{p+2}+\ln\left(\frac{\nu}{\nu_{mm}^{{\text{IC}}}}\right)\right] &\nu_{mm}^{{\mathrm{IC}}}<\nu.
\end{cases}
\end{equation}
(4)$\nu_{c}<\nu_{a}<\nu_{m}$:
\begin{equation}
F_{\nu}^{\text{IC}}=\frac{1}{\Omega}\tau_{\text{es}}F_{\nu,\text{max}}x_0\\
\times\begin{cases}
\left(\frac{1}{2} \mathcal{R}+1\right)(\mathcal{R}+4)\left(\frac{\nu}{\nu_{aa}^{\text{IC}}}\right) &\nu<\nu_{aa}^{\text{IC}},\\
\mathcal{R}\left(\frac{\nu}{\nu_{aa}^{\text{IC}}}\right)^{-\frac{1}{2}}\left[\frac{1}{6} \mathcal{R}+\frac{9}{10}+\frac{1}{4} \mathcal{R} \ln \left(\frac{\nu}{\nu_{a a}^{\text{IC}}}\right)\right] & \nu_{aa}^{\text{IC}}<\nu<\nu_{am}^{\text{IC}},\\
\mathcal{R}^2\left(\frac{\nu}{\nu_{a a}^{\text{IC}}}\right)^{-\frac{1}{2}}\left[\frac{3}{p-1}-\frac{1}{2}+\frac{3}{4} \ln \left(\frac{\nu_{mm}^{\text{IC}}}{\nu}\right)\right] & \nu_{am}^{\text{IC}}<\nu<\nu_{mm}^{\text{IC}},\\
\frac{9}{2(p+2)} \mathcal{R}^2\left(\frac{\nu_a}{\nu_m}\right)\left(\frac{\nu}{\nu_{mm}^{\text{IC}}}\right)^{-\frac{p}{2}}\left[\frac{4}{p+3}\left(\frac{\gamma_a}{\gamma_m}\right)^{p-1} \frac{\gamma_a}{\gamma_c}+\frac{3(p+1)}{(p-1)(p+2)}+\frac{1}{2} \ln \frac{\nu}{\nu_{mm}^{\text{IC}}}\right] & \nu>\nu_{mm}^{\text{IC}}.
\end{cases}
\end{equation}
(4)max$(\nu_{c},\nu_{m})<\nu_{a}$:
\begin{equation}
F_{\nu}^{\text{IC}}=\frac{1}{\Omega}\tau_{\text{es}}F_{\nu,\text{max}}x_0\\
\times\begin{cases}
\left(\frac{3 \mathcal{R}}{2(p+2)}+1\right)\left(\frac{3 \mathcal{R}}{p+2}+4\right)\left(\frac{\nu}{\nu_{aa}^{\text{IC}}}\right) & \nu<\nu_{aa}^{\text{IC}},\\
\frac{1}{p+2}\left[\frac{6 \mathcal{R}}{p+3}+\mathcal{R}\left(\frac{9 \mathcal{R}}{2(p+2)}+1\right)+\frac{9 \mathcal{R}^2}{4} \ln \left(\frac{\nu}{\nu_{aa}^{\text{IC}}}\right)\right]\left(\frac{\nu}{\nu_{aa}^{\text{IC}}}\right)^{-\frac{p}{2}}, & \nu>\nu_{aa}^{\text{IC}}
\end{cases}
\end{equation}
Where $\tau_{\text{es}}$ is the electron scattering optical depth, and $\mathcal{R}$ is the flux ratio between the pile-up peak and the optically thin limit at $\nu_a$.

\section{Klein-Nishina correction}
In higher frequency bands, the SSC is suppressed by the Klein-Nishina effect. Therefore, the shape of spectrum should be corrected for both synchrotron and SSC radiation. \citet{Nakar2009ApJ} have provided a comprehensive treatment of the Klein-Nishina effect. In this section, we refer to their analysis results of the spectrum.

For convenience, the upper limit of the Lorentz factor of electrons that can effectively scatter photons emitted by electrons with a Lorentz factor of $\gamma_e$ is defined as \citep{Nakar2009ApJ}
\begin{equation}
\hat{\gamma}_e=\frac{\gamma m_ec^2}{h\nu_e}.
\end{equation}
Therefore, we can define the new critical Lorentz factors $\hat{\gamma}_m$, $\hat{\gamma}_c$ corresponding to $\nu_m$, $\nu_c$. And since $Y$ is correlated with $\gamma_e$ when the KN effect is significant, a new critical Lorentz factor $\gamma_{e,0}$ can be defined, satisfying $Y(\gamma_{e,0})=1$.

For the fast cooling, the spectrum is divided into three situations based on the relationship between $\gamma_m$ and $\hat{\gamma}_m$. When $\gamma_m<\hat{\gamma}_m$, the spectrum of synchrotron and SCC radiation can be divided into two cases:\\
(1)$\gamma_0<\hat{\gamma}_c$:
\begin{equation}
F_{\nu}^{\text{syn}}\propto\begin{cases}
\nu^{-\frac{1}{2}} &\nu_c<\nu<\nu_m,\\
\nu^{-\frac{p}{2}} &\nu_m<\nu<\hat{\nu}_m,\\
\nu^{-\left(\frac{p}{2}-\frac{1}{4}\right)} &\hat{\nu}_m<\nu<\nu_0,\\
\nu^{-\frac{p}{2}} &\nu_0<\nu.
\end{cases}
\end{equation}
\begin{equation}
F_{\nu}^{\text{IC}}\propto\begin{cases}
\nu^{-\frac{p}{2}} &\nu_{mm}<\nu<2\nu_m\hat{\gamma}_m^2,\\
\nu^{-\left(p-1\right)} &2\nu_m\hat{\gamma}_m^2<\nu<2\nu_m\hat{\gamma}_m\gamma_0,\\
\nu^{-\left(p-\frac{1}{2}\right)} &2\nu_m\hat{\gamma}_m\gamma_0<\nu<2\nu_c\hat{\gamma}_c^2,\\
\nu^{-\left(p+\frac{1}{3}\right)} &2\nu_c\hat{\gamma}_c^2<\nu.
\end{cases}
\end{equation}
(2)$\gamma_0>\hat{\gamma}_c$:
\begin{equation}
F_{\nu}^{\text{syn}}\propto\begin{cases}
\nu^{-\frac{1}{2}} &\nu_c<\nu<\nu_m,\\
\nu^{-\frac{p}{2}} &\nu_m<\nu<\hat{\nu}_m,\\
\nu^{-\left(\frac{p}{2}-\frac{1}{4}\right)} &\hat{\nu}_m<\nu<\hat{\nu}_c,\\
\nu^{-\left(\frac{p}{2}-\frac{2}{3}\right)} &\hat{\nu}_c<\nu<\nu_0,\\
\nu^{-\frac{p}{2}} &\nu_0<\nu.
\end{cases}
\end{equation}
\begin{equation}
F_{\nu}^{\text{IC}}\propto\begin{cases}
\nu^{-\frac{p}{2}} &\nu_m\gamma_m^2<\nu<\nu_m\hat{\gamma}_m^2,\\
\nu^{-\left(p-1\right)} &\nu_m\hat{\gamma}_m^2<\nu<\nu_m\hat{\gamma}_m\gamma_0,\\
\nu^{-\left(p+\frac{1}{3}\right)} &\nu_m\hat{\gamma}_m\gamma_0<\nu.
\end{cases}
\end{equation}
However, when $p\approx2$, another power-law segment is added to the synchrotron spectrum for the case with $\hat{\gamma}_c<\gamma_0$: $F_{\nu}^{\text{syn}}\propto\nu^{-3/8}$ for max$(\nu_c,\hat{\nu_0})<\nu<\hat{\hat{\nu}}_m$.

When $\gamma_m>\hat{\gamma}_m$, the spectrum is divided into three cases. In each case, the spectrum of the SSC radiation is too complex to describe with precise power-law segments \citep{Nakar2009ApJ}. However, the power-law indices of the various segments are typically similar (the difference is $\leq1/4$). Therefore, The spectrum can be described approximately.\\
(1)$\frac{\epsilon_e}{\epsilon_B}<\left(\frac{\gamma_m}{\hat{\gamma}_m}\right)^{\frac{1}{3}}$
\begin{equation}
F_{\nu}^{\text{syn}}\propto\begin{cases}
\nu^{-\frac{1}{2}} &\nu_c<\nu<\hat{\nu}_m,\\
\nu^{-\frac{1}{4}} &\hat{\nu}_m<\nu<\nu_0,\\
\nu^{-\frac{1}{2}} &\nu_0<\nu<\nu_m,\\
\nu^{-\frac{p}{2}} &\nu_m<\nu.
\end{cases}
\end{equation}
For SSC radiation, the peak of $\nu F_{\nu}^{\text{IC}}$ is at $2\nu_m\gamma_m\hat{\gamma}_m$. When the frequencies are much lower than peak frequency, the power-law index of $F_{\nu}^{\text{IC}}$ is $1/4$. At frequencies close to the peak but still below the peak frequency, the index can be between $-3/4$ and $-1/2$. At frequencies are higher than the peak frequency, the power-law index ranges between between $-p+1/4$ and $-p+1/2$, until $\nu>2\nu_c\hat{\gamma}_c^2$, where the power-law index becomes $-p-1/3$.\\
(2)$\left(\frac{\gamma_m}{\hat{\gamma}_m}\right)^{\frac{1}{3}}<\frac{\epsilon_e}{\epsilon_B}<\frac{\gamma_m}{\hat{\gamma}_m}$
\begin{equation}
F_{\nu}^{\text{syn}}\propto\begin{cases}
\nu^{-\frac{1}{2}} &\nu_c<\nu<\hat{\nu}_m,\\
\nu^{-\frac{1}{4}} &\hat{\nu}_m<\nu<\hat{\nu}_0,\\
\nu^{0} &\hat{\nu}_0<\nu<\nu_0,\\
\nu^{-\frac{1}{2}} &\nu_0<\nu<\nu_m,\\
\nu^{-\frac{p}{2}} &\nu_m<\nu.
\end{cases}
\end{equation}
For SSC, at frequencies are just lower than peak frequency ($2\nu_m\gamma_m\hat{\gamma}_m$), the power-law index is $-3/4$. When the frequencies are lower than $\sqrt{\frac{\hat{\gamma}_m\epsilon_e}{\gamma_m\epsilon_B}}2\nu_m\gamma_m\hat{\gamma}_m$, the power-law index becomes $-1/4$. When the frequencies are just higher than peak frequency, the power-law index is between $-p$ and $-p+1/4$. At even higher frequencies, the power-law index is $-p+1/2$. When the frequency reaches $2\nu_c\hat{\gamma}_c^2$ and the power-law index becomes $-p-1/3$.\\
(3)$\frac{\gamma_m}{\hat{\gamma}_m}<\frac{\epsilon_e}{\epsilon_B}<\left(\frac{\gamma_m}{\hat{\gamma}_m}\right)^3$
\begin{equation}
F_{\nu}^{\text{syn}}\propto\begin{cases}
\nu^{-\frac{1}{2}} &\nu_c<\nu<\hat{\nu}_0,\\
\nu^{-\frac{p-1}{4}} &\hat{\nu}_0<\nu<\hat{\nu}_m,\\
\nu^{0} &\hat{\nu}_m<\nu<\nu_m,\\
\nu^{-\frac{p-1}{2}} &\nu_m<\nu<\nu_0,\\
\nu^{-\frac{p}{2}} &\nu_0<\nu.
\end{cases}
\end{equation}
For SSC, the peak frequency of $\nu F_{\nu}^{\text{IC}}$ is $\nu_m\gamma_m\hat{\gamma}_m$. When the frequencies are lower than the peak frequency, the range of power-law index is from $-1/4$ to $0$. And the frequencies are higher than the peak frequency, the range of power-law index is from $-p$ to $-p+1/2$. If the frequencies are higher than $2\nu_c\hat{\gamma}_c^2$, the power-law index is $-p-1/3$.

There is an another case for $\gamma_m=\hat{\gamma}_m$. In this case, the spectrum is divided into 2 cases by $p=2.5$. If $p<2.5$, the synchrotron spectrum is
\begin{equation}
F_{\nu}^{\text{syn}}\propto\begin{cases}
\nu^{-\frac{1}{2}} &\nu_c<\nu<\hat{\nu}_0,\\
\nu^{-\frac{p-1}{3}} &\hat{\nu}_0<\nu<\nu_m,\\
\nu^{-\frac{2\left(p-1\right)}{3}} &\nu_m<\nu<\nu_0,\\
\nu^{-\frac{p}{2}} &\nu_0<\nu.
\end{cases}
\end{equation}
And the SSC spectrum is
\begin{equation}
F_{\nu}^{\text{IC}}\propto\begin{cases}
\nu^{-\frac{1}{2}} &\nu_{cm}<\nu<2\nu_m\hat{\gamma}_0^2,\\
\nu^{-\frac{p-1}{3}} &2\nu_m\hat{\gamma}_0^2<\nu<\nu_{mm},\\
\nu^{-p+1} &\nu_{mm}<\nu<2\nu_m\gamma_m\gamma_0,\\
\nu^{-\frac{2p+1}{3}} &2\nu_m\gamma_m\gamma_0<\nu<2\nu_m\gamma_m\hat{\hat{\gamma}}_0,\\
\nu^{-p+\frac{1}{2}} &2\nu_m\gamma_m\hat{\hat{\gamma}}_0<\nu.
\end{cases}
\end{equation}

When $p>2.5$, for the spectrum of synchrotron, the power-law index between $\nu_c$ and $\nu_m$ is modified to $-1/2$. And at $\nu_m<\nu<\nu_0$, the power-law index is modified to $-p/2+1/4$. For the SSC spectrum, the power-law index between $\nu_{cc}$ and $\nu_{mm}$ is modified to $-1/2$. And at $2\nu_m\gamma_m\gamma_0<\nu$, the power-law index is modified to $-p+1/2$.

For the slow cooling, the synchrotron spectrum above $\nu_c$ is
\begin{equation}
F_{\nu}^{\text{syn}}\propto\begin{cases}
\nu^{-\frac{p}{2}} &\nu_c<\nu<\hat{\nu}_c,\\
\nu^{-\frac{3}{4}\left(p-1\right)} &\text{max}\left(\hat{\nu}_c,\nu_c\right)<\nu<\text{min}\left(\hat{\nu}_m,\nu_0\right),\\
\nu^{-\left(\frac{p}{2}-\frac{2}{3}\right)} &\hat{\nu}_m<\nu<\nu_0,\\
\nu^{-\frac{p}{2}} &\nu_0<\nu.
\end{cases}
\end{equation}
And the spectrum of SSC radiation is
\begin{equation}
F_{\nu}^{\text{IC}}\propto\begin{cases}
\nu^{-\frac{p}{2}} &\nu_{cc}<\nu<2\nu_c\hat{\gamma}_c^2,\\
\nu^{-\left(p-1\right)} &2\nu_c\hat{\gamma}_c\text{max}\left(\hat{\gamma}_c,\gamma_c\right)<\nu<2\nu_c\hat{\gamma}_c\text{min}\left(\gamma_0,\hat{\gamma}_m\right),\\
\nu^{-\frac{p+1}{2}} &2\nu_c\hat{\gamma}_c\gamma_0<\nu<2\nu_c\hat{\gamma}_c\hat{\gamma}_m,\\
\nu^{-\left(p+\frac{1}{3}\right)} &2\nu_c\hat{\gamma}_c\text{max}\left(\gamma_0,\hat{\gamma}_m\right)<\nu.
\end{cases}
\end{equation}

It's worth noting that, for both synchrotron and SSC spectrum, not all these segments can be observed for all cases. For example, the first segment of synchrotron and SSC can be observed only at $\gamma_c<\hat{\gamma}_c$. And the third segment of synchrotron can be observed only at $\hat{\gamma}_m<\gamma_0$. But in this case, the third segment of SSC can't be observed.

\end{document}